\documentclass[aps,amsfonts,amsmath,prd,preprint,nofootinbib,tightenlines]{revtex4-2}

\pdfoutput=1
\newcommand{\beq}{\begin{equation}}
\newcommand{\eeq}{\end{equation}}
\usepackage{graphicx}

\usepackage{xcolor}

\usepackage{tabularx}
\usepackage{multirow}
\usepackage{booktabs}
\usepackage{float}

\usepackage{amsmath}

\newcommand{\be}{\begin{equation}}
\newcommand{\ee}{\end{equation}}
\newcommand{\bea}{\begin{eqnarray}}
\newcommand{\eea}{\end{eqnarray}}
\newcommand{\bi}{\begin{itemize}}
\newcommand{\ei}{\end{itemize}}

\usepackage{overpic,mathtools}

\usepackage{url}
\usepackage[colorlinks]{hyperref}
\hypersetup{ 
	colorlinks=true, 
	linkcolor=black, 
	filecolor=black, 
	citecolor = blue,       
	urlcolor=blue, 
}

\usepackage{tcolorbox}

\begin{document}

\title{Accelerated expansion of the universe purely \\driven by scalar field fluctuations}

\author{Daniel Jim\'enez-Aguilar\footnote{Daniel.Jimenez\_Aguilar@tufts.edu}}

\affiliation{Institute of Cosmology, Department of Physics and Astronomy, Tufts University, Medford, MA 02155, USA\\\\}

\begin{abstract}

We show that scalar field fluctuations alone can drive cosmic acceleration, provided the universe is spatially closed and the Compton wavelength of the field exceeds the radius of curvature. This mechanism may open new perspectives on inflation and dark energy, which could arise from a gas of sufficiently light bosons in a closed universe.   

\end{abstract}

\maketitle

\newpage

\tableofcontents

\newpage

\section{Introduction}

The current accelerated expansion of the universe \cite{SupernovaSearchTeam:1998fmf,SupernovaCosmologyProject:1998vns} is attributed to the presence of a negative-pressure component called dark energy which constitutes, approximately, 69\% of the total energy density in the universe. To date, the nature of this component is unknown. The simplest way to describe it is by means of a cosmological constant $\Lambda$, in which case the dark energy density remains constant in time, leading to the standard $\Lambda$CDM model (``CDM'' stands for ``cold dark matter''). Despite the success of this approach in terms of explaining the observational data, it requires a value of the dark energy density which is in blatant tension with the expected vacuum energy density in the framework of quantum field theory. This is the well-known cosmological constant problem \cite{RevModPhys.61.1}. Moreover, the Dark Energy Spectroscopic Instrument collaboration has recently reported evidence for a time-evolving dark energy density \cite{DESI:2024mwx,DESI:2025zgx}, motivating renewed interest in alternative scenarios such as quintessence models with minimal and nonminimal couplings to gravity \cite{Shlivko:2024llw,Ye:2024ywg,Wolf:2024eph,Wolf:2024stt,Wolf:2025jed,Akrami:2025zlb,Bayat:2025xfr,Cline:2025sbt,Gialamas:2025pwv,Adam:2025kve,SanchezLopez:2025uzw} or dark energy-dark matter interactions \cite{Wang:2024vmw,Li:2024qso,Chakraborty:2025syu,Khoury:2025txd,Guedezounme:2025wav,LaPenna:2026avs,Li:2026xaz}.

Besides the late-time accelerated expansion associated with dark energy, the prevailing cosmological paradigm holds that the very early universe underwent a period of quasi-exponential expansion called inflation \cite{Guth:1980zm,Linde:1981mu,Albrecht:1982wi}. This primeval stage of cosmic acceleration can be driven by a scalar field (the inflaton) that slowly rolls down its potential, solving many of the issues of the standard Big Bang cosmology and providing accurate predictions about the cosmic microwave background (CMB) anisotropies and the large-scale structure in the universe \cite{SDSS:2003eyi,Guth:2005zr,Planck:2018jri}.
Nevertheless, it faces several problems related to the origin and naturalness of the initial state or its robustness against large inhomogeneities in both the scalar field and the metric (see, for instance, \cite{Steinhardt:2011zza,Ijjas:2024oqn,Aurrekoetxea:2024ypv,Hertzberg:2025ifp}). 

The phenomenology of both inflation and dynamical dark energy can be reproduced by means of a homogeneous scalar field that evolves under the influence of a potential. Typically, the shape of this potential and the initial conditions have to be carefully engineered in order to get a successful phase of accelerated expansion. For example, in the case of slow-roll inflation, the potential has to be sufficiently flat and the kinetic energy density of the field must remain subdominant. 
One may argue that this type of constructions is contrived and completely different from the way that other well-known forms of energy in the universe are described. The question motivating this work is whether simple scalar field fluctuations about a minimum of the potential can induce cosmic acceleration. This kind of state seems more natural, as it closely parallels the CMB radiation. In this picture, the accelerated expansion of the universe may simply be a consequence of a ``Higgs boson background radiation'', without the need to invoke a cosmological constant or a specially designed potential. 

One may naively expect that such scalar radiation would exert positive pressure as in the case of the CMB, thereby precluding accelerated expansion. In this work, we will show that this is indeed the case in a spatially flat Friedmann-Lemaître-Robertson-Walker (FLRW) universe, but generally not in a closed one. More specifically, we will consider a well motivated initial state for the field in the form of standard fluctuations in a closed universe and we will prove that the initial equation of state parameter can be smaller than $-1/3$ if the Compton wavelength of the bosons exceeds the initial scale factor. 
As we shall see, this unexpected behavior of radiation is a direct consequence of the non-trivial topology of spacetime. 
The compact topology of the universe plays the role of the physical boundaries in systems where the Casimir effect \cite{Casimir:1948dh} takes place, sourcing additional contributions to the energy density and pressure of the fluctuations \cite{Mamaev:1979ks,Mamaev:1979zw}. 
This has proven relevant in a variety of contexts, in particular in quantum cosmology. For example, back in 1984, Zeldovich and Starobinsky argued that such corrections could have a crucial impact on the quantum creation of universes with non-trivial spatial topologies \cite{Zeldovich:1984vk} (see also \cite{Guth:2025dal} for a recent reexamination of this claim). 

The Casimir effect in cosmological backgrounds has been studied by several authors over the years \cite{PhysRevD.9.341,Streeruwitz:1975wzf,Streeruwitz:1975er,Ford:1975su,Mamaev:1976zb,Ford:1976fn,Dowker:1976pr,Dowker:1978fr,Russell:1987qm,Mostepanenko:1990ceg,Krech1994,Zhuk:1996xc,Elizalde:2003ke,Herdeiro:2005zj,Szydlowski:2007bg,Herdeiro:2009zza,Saharian:2010nep,PhysRevD.83.104042,Bezerra:2011nc,BezerradeMello:2017nyo,Bezerra:2017zqq,Cho:2022ngf,Xie:2023yfg,Hsiang:2024keo,Bezerra:2024wwm,Moreira:2025cwp} (see \cite{Bordag:2001qi} for a review) for scalar, spinor and vector fields. The great majority of this work has focused on conformally coupled fields in the Einstein static universe, with the emphasis placed on the renormalization techniques required to regularize the formally divergent vacuum energy density. The possibility of accelerated expansion was typically not addressed, and in many cases effectively excluded by the assumption of conformal coupling. A notable exception is \cite{Herdeiro:2005zj}, where the authors pointed out that even massless fields minimally coupled to gravity can drive inflation. Although the analysis in that work was restricted to the Einstein static universe, it provided a generalization of earlier results to arbitrary curvature coupling. The present work is similar in spirit to that study in that it emphasizes the dynamical consequences of scalar field fluctuations beyond the conformally coupled case, but it is otherwise completely independent; in particular, all calculations presented here were carried out without prior knowledge of \cite{Herdeiro:2005zj}. A crucial difference is that we derive the initial energy density and pressure of the fluctuations in a non-static universe, as well as the induced Hubble rate and the range of scalar field masses needed for cosmic acceleration. Furthermore, our approach is not restricted to vacuum fluctuations, but instead allows for a generic initial power spectrum. 

The organization of the manuscript is the following. In section \ref{sec:model}, we present the scalar field model under cosnideration. In section \ref{sec:flat universe}, we compute the initial energy density and pressure in a spatially flat FLRW universe for generic curvature coupling and power spectrum of the field, showing that accelerated expansion is impossible. In section \ref{sec:closed universe}, we repeat the calculation for a closed universe and prove that cosmic acceleration is possible in this case. Finally, we present our conclusions in section \ref{sec:conclusions}.


\section{The model}
\label{sec:model}
Consider a universe dominated by a real scalar field $\Phi$ with a vanishing potential energy density at its minimum, located at $\Phi=0$. If the field is sitting at that minimum in the form of small fluctuations, $\Phi(t,\vec{x})=0+\phi(t,\vec{x})$, the action can be expanded as

\begin{equation}
S=\int d^{4}x\sqrt{-g}\left[\frac{R}{16\pi G}-\frac{1}{2}g^{\mu\nu}\partial_{\mu}\phi\partial_{\nu}\phi-V(\phi)-\frac{1}{2}\xi\phi^{2}R\right]\,,
\label{eq:action}
\end{equation}
where 
\begin{equation}
V(\phi)=\frac{1}{2}M^{2}\phi^{2}
\label{eq:potential}
\end{equation}
and $\xi$ is a possible nonminimal coupling to gravity\footnote{As we will see, cosmic acceleration in a closed universe will be possible even for minimal coupling ($\xi=0$).}.
We will assume that these scalar field fluctuations $\phi(t,\vec{x})$ with mass $M$ induce, on average, a FLRW metric:
\begin{equation}
ds^{2}=-dt^{2}+a^{2}(t)\left[d\chi^{2}+\Sigma^{2}(\chi)d\theta^{2}+\Sigma^{2}(\chi)\sin^{2}\theta\,d\varphi^{2}\right]\,,
\label{eq:metric}
\end{equation}
with
\begin{equation}
\Sigma^{2}(\chi)=
\begin{cases}
\sinh^{2}\chi\,\,\,\,\,\,\,\,\,\,\text{if }k=-1\\
\chi^2\,\,\,\,\,\,\,\,\,\,\,\,\,\,\,\,\,\,\,\,\,\,\text{if }k=0\\
\sin^{2}\chi\,\,\,\,\,\,\,\,\,\,\,\,\,\,\text{if }k=1\,.
\end{cases}
\end{equation}
In other words, the metric appearing in the action is not an arbitrary background metric on top of which the fluctuations evolve, but the ensemble-averaged FLRW metric directly generated by them. The universe is completely dominated by the fluctuations. Here and henceforth, we will refer to ensemble-averaged Ricci scalar $R$, Hubble rate $H$, scale factor $a$ and curvature index $k$.

The equation of motion for the scalar field $\phi$ is
\begin{equation}
\ddot{\phi}-\frac{1}{a^{2}}\nabla^{2}\phi+3H\dot{\phi}+\left(M^{2}+\xi R\right)\phi=0\,,
\label{eq:eom phi}
\end{equation}
where dots denote derivatives with respect to cosmic time $t$ and $\nabla^{2}$ is the Laplacian associated with the comoving coordinates $\chi$, $\theta$ and $\varphi$.

The energy-momentum tensor, computed as $T_{\mu\nu}=-(2/\sqrt{-g})\delta S_{\phi}/\delta g^{\mu\nu}$ (with $S_{\phi}$ given by the terms in (\ref{eq:action}) which depend explicitly on $\phi$ and its derivatives), reads
\begin{equation}
T_{\mu\nu}=\partial_{\mu}\phi\partial_{\nu}\phi-g_{\mu\nu}\left[\frac{1}{2}g^{\alpha\beta}\partial_{\alpha}\phi\partial_{\beta}\phi+V(\phi)\right]+\xi\left(G_{\mu\nu}+g_{\mu\nu}\Box-\nabla_{\mu}\nabla_{\nu}\right)\phi^{2}\,,
\label{eq:Tmunu}
\end{equation}
where $G_{\mu\nu}$ is the Einstein tensor, $\nabla_{\mu}$ denotes the covariant derivative and the box operator is defined as $\Box=-\partial_{t}^{2}+(1/a^2)\nabla^{2}-3H\partial_{t}$.
The energy density and pressure, computed as $\rho=T_{00}$ and $p=g^{ij}T_{ij}/3$, respectively, can be shown to be

\begin{equation}
\rho=\frac{1}{2}\dot{\phi}^{2}+\frac{1}{2a^2}(\vec{\nabla}\phi)^{2}+V(\phi)+\xi\left[-\frac{2}{a^2}(\vec{\nabla}\phi)^{2}-\frac{2}{a^2}\phi\nabla^{2}\phi+3\left(H^{2}+\frac{k}{a^2}\right)\phi^{2}+6H\phi\dot{\phi}\right]\,,
\label{eq:energy density phi}
\end{equation}

\begin{equation}
\begin{split}
p&=\frac{1}{2}\dot{\phi}^{2}-\frac{1}{6a^2}(\vec{\nabla}\phi)^{2}-V(\phi)+\\
&+\xi\left\{-2\dot{\phi}^{2}+\frac{4}{3a^2}(\vec{\nabla}\phi)^{2}-\frac{2}{3a^2}\phi\nabla^{2}\phi+2\phi V'(\phi)+\left[H^{2}+\frac{k}{a^2}+\left(2\xi-\frac{1}{3}\right)R\right]\phi^{2}+2H\phi\dot{\phi}\right\}\,.
\label{eq:pressure phi}
\end{split}
\end{equation}

If we introduce conformal time (defined as $d\eta=dt/a$) and the rescaled field
\begin{equation}
\psi(\eta,\vec{x})=a(\eta)\phi(\eta,\vec{x})\,,
\label{eq:chi}
\end{equation}
the equation of motion (\ref{eq:eom phi}) can be rewritten as
\begin{equation}
\psi''-\nabla^{2}\psi+a^{2}M_{\rm eff}^{2}\,\psi=0\,,
\label{eq:eom chi}
\end{equation}
where primes indicate derivatives with respect to conformal time $\eta$ and the time-dependent effective mass $M_{\rm eff}(\eta)$ is given by
\begin{equation}
M_{\rm eff}^{2}(\eta)=M^{2}+\left(\xi-\frac{1}{6}\right)R(\eta)+\frac{k}{a^2(\eta)}\,.
\label{eq:effective mass}
\end{equation}
In terms of the field $\psi$, the energy density and pressure become

\begin{equation}
\begin{split}
\rho&=\frac{1}{2a^{4}}(\psi')^2+\frac{1-4\xi}{2a^4}(\vec{\nabla}\psi)^2+\frac{1}{2a^2}\left[M^2+(1-6\xi)H^2+6\xi\frac{k}{a^2}\right]\psi^2-\frac{2\xi}{a^4}\psi\nabla^{2}\psi-\frac{H}{a^3}(1-6\xi)\psi\psi'\,,
\label{eq:energy density chi}
\end{split}
\end{equation}

\begin{equation}
\begin{split}
p&=\frac{1-4\xi}{2a^{4}}(\psi')^2-\frac{1-8\xi}{6a^4}(\vec{\nabla}\psi)^2+\frac{1}{2a^2}\left[(4\xi-1)M^2+(1-6\xi)H^2+2\xi\frac{k}{a^2}+\left(4\xi-\frac{2}{3}\right)\xi R\right]\psi^2-\\
&-\frac{2\xi}{3a^4}\psi\nabla^{2}\psi-\frac{H}{a^3}(1-6\xi)\psi\psi'\,.
\label{eq:pressure chi}
\end{split}
\end{equation}
Given the initial conditions for $\psi$, and in particular its power spectrum, one can use the last two expressions to compute the ensemble-averaged energy density and pressure and determine whether the universe is accelerating at the initial time. According to the acceleration equation, we will have
\begin{equation}
\frac{\ddot{a}_{i}}{a_{i}}=-\frac{4\pi G}{3}(\langle\rho_{i}\rangle+3\langle p_{i}\rangle)\,,
\end{equation}
where the subscript $i$ denotes evaluation at the initial time. In the following sections, we will determine whether $\langle\rho_i\rangle+3\langle p_i\rangle<0$, in which case the universe will undergo accelerated expansion.


\section{Spatially flat universe}
\label{sec:flat universe}
For $k=0$, we self-consistently choose the initial conditions
\begin{equation}
\psi(\eta_{i},\vec{x})=\int\frac{d^{3}q}{\sqrt{2(2\pi)^{3}\omega_{\vec{q}}}}\left(\alpha_{\vec{q}}e^{i\vec{q}\cdot\vec{x}}+\alpha_{\vec{q}}^{*}e^{-i\vec{q}\cdot\vec{x}}\right)\,,
\label{eq:expansion chi}
\end{equation}
\begin{equation}
\psi'(\eta_{i},\vec{x})=-i\int d^{3}q\sqrt{\frac{\omega_{\vec{q}}}{2(2\pi)^{3}}}\left(\alpha_{\vec{q}}e^{i\vec{q}\cdot\vec{x}}-\alpha_{\vec{q}}^{*}e^{-i\vec{q}\cdot\vec{x}}\right)\,,
\label{eq:expansion chi prime}
\end{equation}
which are well motivated by the equation of motion (\ref{eq:eom chi}). Indeed, at the quantum level, these conditions minimize the expectation value of the Hamiltonian in the initial vacuum state (see, for example, chapter 6 in \cite{Mukhanov:2007zz}). In these expressions, the comoving angular frequency is given by
\begin{equation}
\omega_{\vec{q}}^{2}=\vec{q}^{\,\,2}+a_{i}^{2}M_{\rm eff}^{2}(\eta_{i})\,.
\label{eq:omega}
\end{equation}
On the other hand, the real and imaginary parts of the complex coefficients $\alpha_{\vec{q}}$ will be taken to be independent random variables with zero mean and variance $\langle|\alpha_{\vec{q}}|^{2}\rangle/2$. One can use this information to find the initial ensemble averages of $\psi^2$, $(\psi')^2$, $(\vec{\nabla}\psi)^2$, $\psi\nabla^{2}\psi$ and $\psi\psi'$, which are needed to compute the averages of (\ref{eq:energy density chi}) and (\ref{eq:pressure chi}):
\begin{equation}
\langle\psi^{2}\rangle=\frac{1}{(2\pi)^3}\int d^{3}q\,\frac{\langle|\alpha_{\vec{q}}|^{2}\rangle}{\omega_{\vec{q}}}\,,
\label{eq:mean 1}
\end{equation}
\begin{equation}
\langle(\psi')^{2}\rangle=\frac{1}{(2\pi)^3}\int d^{3}q\,\omega_{\vec{q}}\langle|\alpha_{\vec{q}}|^{2}\rangle\,,
\label{eq:mean 2}
\end{equation}
\begin{equation}
\langle(\vec{\nabla}\psi)^2\rangle=\frac{1}{(2\pi)^3}\int d^{3}q\,\frac{\vec{q}^{\,\,2}\langle|\alpha_{\vec{q}}|^{2}\rangle}{\omega_{\vec{q}}}=-\langle\psi\nabla^{2}\psi\rangle\,,
\label{eq:mean 3}
\end{equation}
\begin{equation}
\langle\psi\psi'\rangle=0\,.
\label{eq:mean 4}
\end{equation}
The reader is referred to the appendix for details on the derivation of these expressions. Substituting them into (\ref{eq:energy density chi}) and (\ref{eq:pressure chi}), we get 
\begin{equation}
\langle\rho_{i}\rangle=\frac{1}{(2\pi)^{3}a_{i}^{4}}\int d^{3}q\,\omega_{\vec{q}}\,\langle|\alpha_{\vec{q}}|^{2}\rangle+\frac{1}{2(2\pi)^{3}a_{i}^{2}}(1-6\xi)\left(H_{i}^{2}+\frac{R_{i}}{6}\right)\int d^{3}q\,\frac{\langle|\alpha_{\vec{q}}|^{2}\rangle}{\omega_{\vec{q}}}\,,
\label{eq:ensemble average rho}
\end{equation}
\begin{equation}
\langle p_{i}\rangle=\frac{1}{3(2\pi)^{3}a_{i}^{4}}\int d^{3}q\,\frac{\vec{q}^{\,\,2}\langle|\alpha_{\vec{q}}|^{2}\rangle}{\omega_{\vec{q}}}+\frac{1}{2(2\pi)^{3}a_{i}^{2}}(1-6\xi)\left(H_{i}^{2}-\frac{R_{i}}{6}\right)\int d^{3}q\,\frac{\langle|\alpha_{\vec{q}}|^{2}\rangle}{\omega_{\vec{q}}}\,.
\label{eq:ensemble average pressure}
\end{equation}
\\
For standard vacuum fluctuations we have 
$\langle|\alpha_{\vec{q}}|^{2}\rangle=1/2$,
and for thermal fluctuations at temperature $T$,
\begin{equation}
\langle|\alpha_{\vec{q}}|^{2}\rangle=\frac{1}{e^{\omega_{\vec{q}}/\tilde{T}}-1}\,,
\label{eq:occupation number thermal fluctuations}
\end{equation}
with $\tilde{T}=a_{i}T$. In any case, it is clear that the usual expressions for Minkowski space are recovered if the second terms on the right-hand sides vanish. This happens in a static universe ($H_{i}=0$, $R_{i}=0$) or if the field is conformally coupled to gravity ($\xi=1/6$). Otherwise, corrections to those expressions exist in an expanding universe, even for minimal coupling ($\xi=0$). Can the second term in (\ref{eq:ensemble average pressure}) make the pressure sufficiently negative to cause accelerated expansion? To answer this question, note first that the second terms on the right-hand sides depend implicitly on the left-hand sides. Using
\begin{equation}
H_{i}^{2}+\frac{R_{i}}{6}=4\pi G(\langle\rho_{i}\rangle-\langle p_{i}\rangle)
\label{eq:simplification 1}
\end{equation}
and
\begin{equation}
H_{i}^{2}-\frac{R_{i}}{6}=
\frac{4\pi G}{3}(\langle\rho_{i}\rangle+3\langle p_{i}\rangle)\,,
\label{eq:simplification 2}
\end{equation}
and defining the integrals
\begin{equation}
\rho_{0}=\frac{1}{(2\pi)^{3}a_{i}^{4}}\int d^{3}q\,\omega_{\vec{q}}\,\langle|\alpha_{\vec{q}}|^{2}\rangle\,,
\label{eq:rho 0}
\end{equation}

\begin{equation}
p_{0}=\frac{1}{3(2\pi)^{3}a_{i}^{4}}\int d^{3}q\,\frac{\vec{q}^{\,\,2}\langle|\alpha_{\vec{q}}|^{2}\rangle}{\omega_{\vec{q}}}\,,
\label{eq:p 0}
\end{equation}

\begin{equation}
I=\frac{1}{(2\pi)^{3}a_{i}^{2}}\int d^{3}q\,\frac{\langle|\alpha_{\vec{q}}|^{2}\rangle}{\omega_{\vec{q}}}\,,
\label{eq:I}
\end{equation}

\begin{equation}
x=2\pi IG(1-6\xi)\,,
\label{eq:x}
\end{equation}
one can solve for $\langle\rho_{i}\rangle$ and $\langle p_{i}\rangle$ in (\ref{eq:ensemble average rho}) and (\ref{eq:ensemble average pressure}) and rewrite them in the compact form
\vspace{0.5\baselineskip}
\begin{tcolorbox}
\begin{equation}
\langle\rho_{i}\rangle=\frac{3(1-x)\rho_{0}-3xp_{0}}{4x^{2}-6x+3}\,,
\label{eq:ensemble average rho final}
\end{equation}

\begin{equation}
\langle p_{i}\rangle=\frac{3(1-x)p_{0}+x\rho_{0}}{4x^{2}-6x+3}\,.
\label{eq:ensemble average p final}
\end{equation}
\end{tcolorbox}
\vspace{0.25\baselineskip}
One can finally use these two expressions to show that cosmic acceleration cannot occur. Taking into account that the polynomial $4x^{2}-6x+3$ is positive for any $x$, we first note that $\langle\rho_{i}\rangle+3\langle p_{i}\rangle<0$ requires
\begin{equation}
\rho_{0}<(4x-3)p_{0}\,.
\label{eq:condition flat one}
\end{equation}
However, since we are considering a spatially flat universe, Friedmann's equation imposes the additional condition $\langle\rho_{i}\rangle>0$, which implies
\begin{equation}
(1-x)\rho_{0}>xp_{0}\,.
\label{eq:condition flat two}
\end{equation}
Since both $\rho_{0}$ and $p_{0}$ are strictly positive, the last inequality can only be satisfied for $x<1$. Otherwise, the left-hand side would be negative and the right-hand side would be positive. Now, if we multiply both sides of (\ref{eq:condition flat one}) by $1-x$, we get
\begin{equation}
(1-x)\rho_{0}<(-4x^{2}+7x-3)p_{0}=-(4x^{2}-6x+3)p_{0}+xp_{0}\,.
\end{equation}
Since $4x^{2}-6x+3$ is positive, the right-hand side is smaller than $xp_{0}$. Therefore, (\ref{eq:condition flat one}) and (\ref{eq:condition flat two}) are clearly incompatible, meaning that $\langle\rho_{i}\rangle+3\langle p_{i}\rangle<0$ and $\langle\rho_{i}\rangle>0$ cannot be satisfied simultaneously. Note that the incompatibility holds for any curvature coupling $\xi$ and any choice of spectrum $\langle|\alpha_{\vec{q}}|^{2}\rangle$. 

It should be stressed that this proof relies on the fact that $\rho_{0}$ and $p_{0}$ are positive, as follows from (\ref{eq:rho 0}) and (\ref{eq:p 0}). However, this property may not hold for quantum vacuum fluctuations. In this case, the aforementioned expressions are formally divergent and one has to apply a renormalization procedure, after which the resulting finite quantities are not guaranteed to remain positive. Therefore, the problem of quantum vacuum fluctuations requires additional care and is left for future work. 

\begin{figure}[h!]
\centering
\includegraphics[scale=0.42]{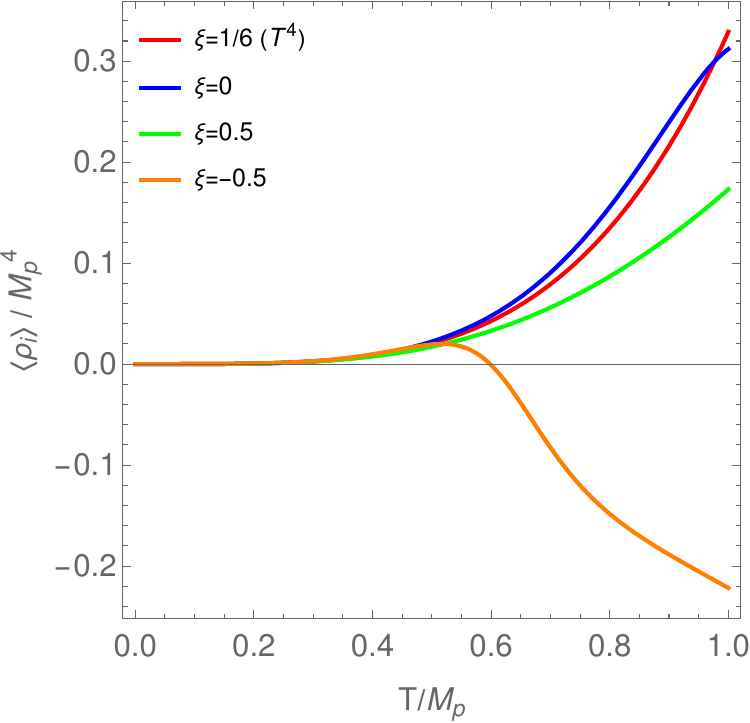}
\includegraphics[scale=0.42]{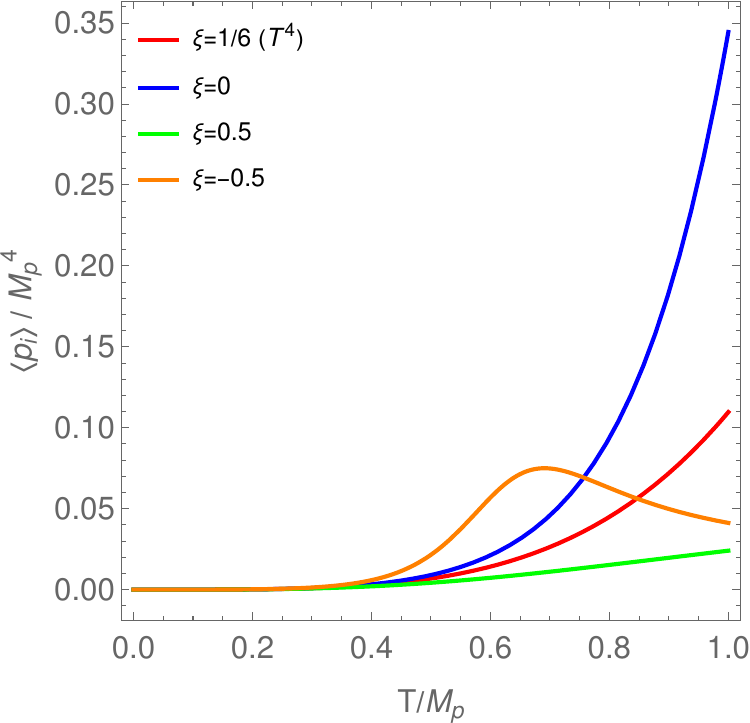}
\includegraphics[scale=0.412]{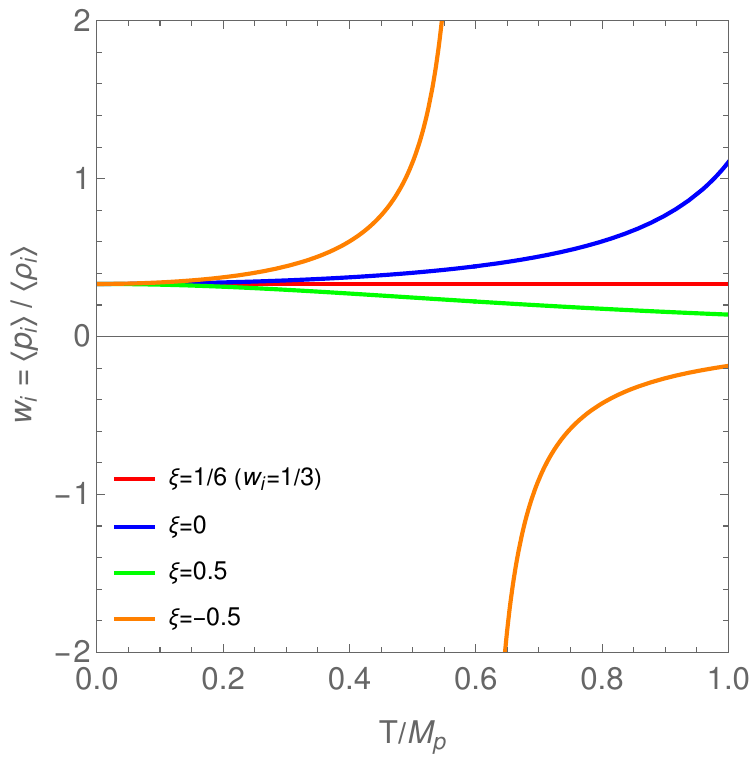}
\caption{Thermal energy density (left), pressure (center) and equation of state parameter $w_{i}=\langle p_{i}\rangle/\langle\rho_{i}\rangle$ (right) in the high-temperature limit for several values of the nonminimal coupling parameter $\xi$. The usual $\langle\rho\rangle\propto T^{4}$ law (with $w=1/3$) is recovered for the conformal coupling $\xi=1/6$. For other values of $\xi$ (even for $\xi=0$, which corresponds to the field being minimally coupled to gravity), this law gets corrections.}
\label{fig:rho p w vs T}
\end{figure}

Even if accelerated expansion is not possible in the flat universe (with the caveat above), it is interesting to see how the usual $\langle\rho\rangle\propto T^{4}$ law in the case of a thermal spectrum is modified in the high-temperature limit, defined as $T\gg M_{\rm eff}(\eta_{i})$. This is shown in figure \ref{fig:rho p w vs T}.
In this limit, for $\xi=0$, we have $x\sim(T/M_{p})^2$, with $M_{p}$ the Planck mass. If $x\ll 1$ ($T\ll M_{p}$), one can easily check from (\ref{eq:ensemble average rho final}) and (\ref{eq:ensemble average p final}) that the deviations from the standard $\rho_{0}$ and $p_{0}$ are very small. However, at temperatures closer to the Planck mass, the corrections are significant.

As a final comment, let us note that the results shown here are valid under the assumption of small scalar field fluctuations (that is, the assumption of a quadratic potential), which may or may not be consistent with Planckian temperatures depending on the mass of the field and the curvature of the full potential at its minimum.
Let $W(\Phi)$ be the full potential and consider the next few terms in the Taylor expansion around the minimum at $\Phi=0$:
\begin{equation}
W(\phi)\approx\frac{1}{2}M^{2}\phi^{2}+\frac{1}{6}A\phi^{3}+\frac{1}{24}B\phi^{4}\,,
\end{equation}
where $A$ and $B$ are the third and fourth derivatives of the potential at that point, respectively. On average, the second term on the right-hand side vanishes, so the quadratic approximation will break down when the the ensemble average of the third term becomes comparable to that of the first term. In other words, the quadratic approximation should remain valid for temperatures such that
\begin{equation}
M^{2}\langle\phi^{2}\rangle\gg\frac{|B|}{12}\langle\phi^{4}\rangle\,.
\end{equation} 
In the high-temperature limit, we have $\langle\phi^{2}\rangle\propto T^{2}$ and $\langle\phi^{4}\rangle\propto T^{4}$. Dropping numerical prefactors, we conclude that 
\begin{equation}
T\ll\frac{M}{\sqrt{|B|}}
\end{equation}
is needed in order for the quadratic approximation to hold. Therefore, Planckian temperatures are consistent with ``small fluctuations'' of the scalar field if 
\begin{equation}
|B|\ll\left(\frac{M}{M_{p}}\right)^{2}\,.
\end{equation}


\section{Spatially closed universe}
\label{sec:closed universe}
In the case $k=1$, we can choose initial conditions analogous to the ones used in the previous section by expanding the field $\psi$ in terms of hyperspherical harmonics \cite{PhysRevD.9.341} (see also chapter 5 in \cite{Birrell:1982ix}):
\begin{equation}
\psi(\eta_{i},\vec{x})=\sum_{J}\frac{1}{\sqrt{2\omega_{n}}}\left[\alpha_{n}Y_{_J}(\vec{x})+\alpha_{n}^{*}Y_{_J}^{*}(\vec{x})\right]\,,
\label{eq:expansion psi closed}
\end{equation}
\begin{equation}
\psi'(\eta_{i},\vec{x})=-i\sum_{J}\sqrt{\frac{\omega_{n}}{2}}\left[\alpha_{n}Y_{_J}(\vec{x})-\alpha_{n}^{*}Y_{_J}^{*}(\vec{x})\right]\,,
\label{eq:expansion psi prime closed}
\end{equation}
where 
\begin{equation}
\omega_{n}^{2}=n(n+2)+a_{i}^{2}M_{\rm eff}^{2}(\eta_{i})
\label{eq:omega closed}
\end{equation}
and the index $J$ is used for the three discrete indices $n=0,1,...$; $l=0,1,...,n$, and $m=-l,-l+1,...,l$. In a closed universe, the hyperspherical harmonic $Y_{_J}(\vec{x})$ is a solution of
\begin{equation}
\nabla^{2}Y_{_J}(\vec{x})=-n(n+2)Y_{_J}(\vec{x})
\label{eq:hyperspherical harmonic equation}
\end{equation}
and is given by
\begin{equation}
Y_{_J}(\vec{x})=\Pi^{(+)}_{nl}(\chi)\,y_{lm}(\theta,\varphi)\,,
\label{eq:YJ}
\end{equation}
with $y_{lm}(\theta,\varphi)$ the usual spherical harmonics and 
\begin{equation}
\Pi^{(+)}_{nl}(\chi)=\left[\frac{\pi n^{2}(1-n^{2})\cdot\cdot\cdot(l^{2}-n^{2})}{2}\right]^{-1/2}\left(i\sin\chi\right)^{l}\frac{d^{1+l}\cos(n\chi)}{d(\cos\chi)^{1+l}}\,,
\end{equation}
which can also be expressed in terms of Gegenbauer polynomials \cite{Ford:1976fn}.

Proceeding as in the previous section, one finds
\begin{equation}
\langle\psi^{2}\rangle=\frac{1}{2\pi^2}\sum_{n=0}^{\infty}\frac{(n+1)^{2}\langle|\alpha_{n}|^{2}\rangle}{\omega_{n}}\,,
\label{eq:mean 1 closed}
\end{equation}
\begin{equation}
\langle(\psi')^{2}\rangle=\frac{1}{2\pi^2}\sum_{n=0}^{\infty}(n+1)^{2}\omega_{n}\langle|\alpha_{n}|^{2}\rangle\,,
\label{eq:mean 2 closed}
\end{equation}
\begin{equation}
\langle(\vec{\nabla}\psi)^2\rangle=\frac{1}{2\pi^2}\sum_{n=0}^{\infty}\frac{n(n+2)(n+1)^{2}\langle|\alpha_{n}|^{2}\rangle}{\omega_{n}}=-\langle\psi\nabla^{2}\psi\rangle\,,
\label{eq:mean 3 closed}
\end{equation}
\begin{equation}
\langle\psi\psi'\rangle=0\,.
\label{eq:mean 4 closed}
\end{equation}
Substituting (\ref{eq:mean 1 closed})-(\ref{eq:mean 4 closed}) into (\ref{eq:energy density chi}) and (\ref{eq:pressure chi}), we get
\begin{equation}
\langle\rho_{i}\rangle=\frac{1}{2\pi^2 a_{i}^{4}}\sum_{n=0}^{\infty}(n+1)^{2}\omega_{n}\langle|\alpha_{n}|^{2}\rangle+\frac{1-6\xi}{4\pi^{2}a_{i}^{2}}\left(H_{i}^{2}+\frac{R_{i}}{6}-\frac{1}{a_{i}^2}\right)\sum_{n=0}^{\infty}\frac{(n+1)^{2}\langle|\alpha_{n}|^{2}\rangle}{\omega_{n}}\,,
\label{eq:rho closed}
\end{equation}

\begin{equation}
\begin{split}
\langle p_{i}\rangle&=\frac{1}{2\pi^2 a_{i}^{4}}\sum_{n=0}^{\infty}\frac{n(n+2)(n+1)^{2}\langle|\alpha_{n}|^{2}\rangle}{3\omega_{n}}+\frac{1}{4\pi^{2}a_{i}^{2}}\left[(1-6\xi)\left(H_{i}^{2}-\frac{R_{i}}{6}\right)+\frac{1-2\xi}{a_{i}^2}\right]\sum_{n=0}^{\infty}\frac{(n+1)^{2}\langle|\alpha_{n}|^{2}\rangle}{\omega_{n}}\,.
\label{eq:pressure closed}
\end{split}
\end{equation}
Now we use 
\begin{equation}
H_{i}^{2}+\frac{R_{i}}{6}=4\pi G(\langle\rho_{i}\rangle-\langle p_{i}\rangle)-\frac{1}{a_{i}^2}
\label{eq:simplification 1 closed}
\end{equation}
and
\begin{equation}
H_{i}^{2}-\frac{R_{i}}{6}=
\frac{4\pi G}{3}(\langle\rho_{i}\rangle+3\langle p_{i}\rangle)-\frac{1}{a_{i}^2}\,,
\label{eq:simplification 2 closed}
\end{equation}
and define
\begin{equation}
\rho_{0}=\frac{1}{2\pi^2 a_{i}^{4}}\sum_{n=0}^{\infty}(n+1)^{2}\omega_{n}\langle|\alpha_{n}|^{2}\rangle\,,
\label{eq:rho 0 closed}
\end{equation}

\begin{equation}
p_{0}=\frac{1}{2\pi^2 a_{i}^{4}}\sum_{n=0}^{\infty}\frac{n(n+2)(n+1)^{2}\langle|\alpha_{n}|^{2}\rangle}{3\omega_{n}}\,,
\label{eq:p 0 closed}
\end{equation}

\begin{equation}
I=\frac{1}{2\pi^{2}a_{i}^{2}}\sum_{n=0}^{\infty}\frac{(n+1)^{2}\langle|\alpha_{n}|^{2}\rangle}{\omega_{n}}\,,
\label{eq:I closed}
\end{equation}

\begin{equation}
x=2\pi IG(1-6\xi)\,.
\end{equation}
Inserting these expressions into (\ref{eq:rho closed}) and (\ref{eq:pressure closed}) and solving for $\langle\rho_{i}\rangle$ and $\langle p_{i}\rangle$, we find
\vspace{0.5\baselineskip}
\begin{tcolorbox}
\begin{equation}
\langle\rho_{i}\rangle=\frac{3(1-x)\rho_{0}-3xp_{0}}{4x^{2}-6x+3}+\frac{3I}{a_{i}^2}\,\frac{(1-8\xi)x+6\xi-1}{4x^{2}-6x+3}\,,
\label{eq:ensemble average rho final closed}
\end{equation}

\begin{equation}
\langle p_{i}\rangle=\frac{3(1-x)p_{0}+x\rho_{0}}{4x^{2}-6x+3}-\frac{I}{a_{i}^2}\,\frac{x-6\xi}{4x^{2}-6x+3}\,.
\label{eq:ensemble average p final closed}
\end{equation}
\end{tcolorbox}
\vspace{0.25\baselineskip}
Due to the compact topology of spacetime, note that the ensemble-averaged energy density and pressure in a spatially closed universe get an additional term with respect to the flat case. Note also that the extra term in (\ref{eq:ensemble average p final closed}) can contribute negatively to the pressure, in particular for minimal coupling ($\xi=0$).

\subsection{Sanity check: conformal coupling}
As a sanity check, we may consider the case of conformal coupling, which is well known in the literature. Setting $\xi=1/6$ ($x=0$), one gets 
\begin{equation}
\langle\rho_i\rangle=\rho_{0}
\end{equation}
and
\begin{equation}
\langle p_i\rangle=p_{0}+\frac{I}{3a_{i}^2}=\frac{1}{6\pi^{2}a_{i}^4}\sum_{n=0}^{\infty}\frac{(n+1)^{2}\langle|\alpha_{n}|^{2}\rangle}{\omega_{n}}\left(\omega_{n}^2-a_{i}^{2}M^2\right)\,,
\end{equation}
where in the last line we have used $\omega_{n}^2=(n+1)^{2}+a_{i}^{2}M^2$, which can be easily shown using (\ref{eq:effective mass}) and (\ref{eq:omega closed}). These results agree with the ones reported in \cite{Ford:1976fn} for vacuum fluctuations. It should be emphasized that the pressure is manifestly positive in this case, so fluctuations of a conformally coupled scalar cannot produce cosmic acceleration. We also note that the equation of state parameter is 1/3 if the field is massless, as expected.

\subsection{Accelerated expansion: minimal coupling}
Now we will show that the following two conditions can be satisfied simultaneously for other couplings:
\\\\
(i). $\langle\rho_{i}\rangle+3\langle p_{i}\rangle<0$.\\\\
(ii). $H_{i}^{2}=\frac{8\pi G}{3}\langle\rho_{i}\rangle-\frac{1}{a_{i}^{2}}>0$.\\\\
This will imply that a spatially closed universe dominated by scalar field fluctuations can undergo accelerated expansion. In order to exemplify this, let us focus on the minimal coupling case $\xi=0$ for simplicity. From (\ref{eq:ensemble average rho final closed}) and (\ref{eq:ensemble average p final closed}) it follows that $\langle\rho_{i}\rangle+3\langle p_{i}\rangle<0$ if
\begin{equation}
\tilde{I}>\frac{\tilde{\rho}_{0}+3\tilde{p}_{0}}{1+\frac{8\pi}{\tilde{a}_{i}^{2}}\tilde{p}_{0}}\,,
\label{eq:condition accelerated expansion}
\end{equation}
where we have defined the dimensionless variables $\tilde{I}=a_{i}^{2}I$, $\tilde{\rho}_{0}=a_{i}^{4}\rho_{0}$, $\tilde{p}_{0}=a_{i}^{4} p_{0}$ and $\tilde{a}_{i}=a_{i}/L_{p}$, with $L_{p}$ the Planck length. Note that
\begin{equation}
\tilde{I}=\frac{\langle|\alpha_{0}|^{2}\rangle}{2\pi^{2}\omega_{0}}+\frac{1}{2\pi^{2}}\sum_{n=1}^{\infty}\frac{(n+1)^{2}\langle|\alpha_{n}|^{2}\rangle}{\omega_{n}}\,,
\label{eq:I tilde}
\end{equation}
so (\ref{eq:condition accelerated expansion}) can be easily satisfied if $\omega_{0}$ is sufficiently small. Indeed, in the limit $\omega_{0}\rightarrow 0$, we see that $\tilde{I}\rightarrow\infty$ while $\tilde{\rho}_{0}$ and $\tilde{p}_{0}$ approach finite values. An important assumption we are making here is that the infinite sums are convergent and positive. This naturally happens for a thermal spectrum of the form (\ref{eq:occupation number thermal fluctuations}), but not for 
vacuum fluctuations\footnote{Similarly to the spatially flat case, the sums (\ref{eq:rho 0 closed}), (\ref{eq:p 0 closed}) and (\ref{eq:I closed}) may not remain positive after the renormalization procedure and the conclusions extracted from this analysis could change.}.  
With this in mind, let us further assume that $\omega_{0}$ is so small that $\tilde{I}\gg\tilde{\rho}_{0},\tilde{p}_{0}$. In this case, condition (i) above, or, equivalently, equation (\ref{eq:condition accelerated expansion}), is clearly satisfied. In order for condition (ii) to be satisfied as well, one can use (\ref{eq:ensemble average rho final closed}) to show that $x>3/2$ is needed. In terms of the dimensionless variables, $x$ is given by 
\begin{equation}
x=\frac{2\pi\tilde{I}}{\tilde{a}_{i}^{2}}\,.
\label{eq:x dimensionless}
\end{equation} 
Therefore, even if $\tilde{I}$ is huge, $x$ can be order 1 if $\tilde{a}_{i}$ is big. The range $x\in\left(3/2,\infty\right)$ can be translated into a range for $\omega_{0}$: $\omega_{0}\in\left(0,\Omega_{0}\right)$. For instance, consider the thermal spectrum (\ref{eq:occupation number thermal fluctuations}), and suppose that (\ref{eq:I tilde}) is dominated by the first term on the right-hand side. Assume also that we choose $\tilde{T}\gg\omega_{0}$, so that $\langle|\alpha_{0}|^{2}\rangle\approx\tilde{T}/\omega_{0}$. Then, using (\ref{eq:x dimensionless}), we see that $x>3/2$ means
\begin{equation}
\omega_{0}<\sqrt{\frac{2\tilde{T}}{3\pi\tilde{a}_{i}^{2}}}\equiv\Omega_{0}\,.
\end{equation} 
Once $\tilde{a}_{i}$ and the parameters characterizing the spectrum ($\tilde{T}$ in the thermal case) are chosen, we can find a continuum of configurations that allow for accelerated expansion as we vary $\omega_{0}$ from 0 to $\Omega_{0}$. A concrete example is the following: if the initial scale factor is $\tilde{a}_{i}=100$, for a thermal state at temperature $\tilde{T}=0.1$, $\omega_{0}$ in the range $(0,0.015)$ produces accelerated expansion (and the Hubble rate is real). For instance, for $\omega_{0}=0.0012$, we get the equation of state parameter $w_{i}=\langle p_{i}\rangle/\langle\rho_{i}\rangle\approx-0.61$ and the Hubble rate $H_{i}\approx\pm 0.0039M_{p}$.

A question that may arise at this point is the following. We have
\begin{equation}
\omega_{0}^{2}=a_{i}^{2}\left(M^{2}-\frac{R_{i}}{6}+\frac{1}{a_{i}^2}\right)\,,
\label{eq:omega 0}
\end{equation}
where 
\begin{equation}
R_{i}=8\pi G\left(\langle\rho_{i}\rangle-3\langle p_{i}\rangle\right)
\label{eq:Ricci scalar} 
\end{equation}
is the initial Ricci scalar. Consider the specific example indicated above. Clearly, $R_{i}$ depends implicitly on temperature, so $\omega_{0}$ and $T$ are not independent variables. How do we know that $\omega_{0}=0.0012$ is compatible with $T=\tilde{T}/a_{i}=10^{-3}M_{p}$? Since $\omega_{0}$ also depends on the bare mass of the scalar field, $M$, one can always make $\omega_{0}$ and $T$ compatible by choosing $M$ appropriately.

In conclusion, accelerated expansion can be achieved for a specific range of scalar field masses. This range, as well as the corresponding Hubble rate and equation of state parameter, can be easily deduced when $\tilde{I}\gg\tilde{\rho}_{0},\tilde{p}_{0}$. In this limit, (\ref{eq:ensemble average rho final closed}) and (\ref{eq:ensemble average p final closed}) reduce to
\begin{equation}
\langle\rho_{i}\rangle\approx\frac{3x(x-1)}{2\pi Ga_{i}^{2}\left(4x^{2}-6x+3\right)}\,,
\label{eq:rho i limit}
\end{equation}
\begin{equation}
\langle p_{i}\rangle\approx-\frac{x^2}{2\pi Ga_{i}^{2}\left(4x^{2}-6x+3\right)}\,.
\label{eq:p i limit}
\end{equation}
\\
It immediately follows that the equation of state parameter is
\begin{equation}
w_{i}\approx-\frac{x}{3\left(x-1\right)}\,.
\label{eq:w i limit}
\end{equation}
If $x\rightarrow\infty$, then $w_{i}\rightarrow-1/3$, and if $x\rightarrow 3/2$, $w_{i}\rightarrow-1$. Now, combining (\ref{eq:rho i limit}) with Friedmann's equation, we get
\begin{equation}
H_{i}\approx\pm\frac{1}{a_{i}}\sqrt{\frac{2x-3}{4x^{2}-6x+3}}\,.
\label{eq:H i limit}
\end{equation}
Note that $H_{i}\rightarrow 0$ in the two limiting cases $x\rightarrow\infty$ and $x\rightarrow 3/2$. However, $\dot{H}_{i}$ is different in each case. Inserting (\ref{eq:rho i limit}) and (\ref{eq:p i limit}) into
\begin{equation}
\dot{H}_{i}=-4\pi G\left(\langle p_{i}\rangle+\langle\rho_{i}\rangle\right)+\frac{1}{a_{i}^2}\,,
\end{equation}
we find
\begin{equation}
\dot{H}_{i}\approx\frac{3}{a_{i}^{2}(4x^{2}-6x+3)}\,.
\end{equation}
\\
This implies that $\dot{H}_{i}\rightarrow 0$ as $x\rightarrow\infty$, but $\dot{H}_{i}\rightarrow 1/a_{i}^2$ as $x\rightarrow 3/2$. Therefore, we may refer to $x\rightarrow\infty$ as the ``static universe limit''. In the other limit, $x\rightarrow 3/2$, the universe will start out with $H_{i}=0$ and $w_{i}=-1$, and will subsequently inflate. This is reminiscent of a bounce or a quantum-driven onset of inflation \cite{Hartle:1983ai,Vilenkin:1984wp}.

Finally, we can insert (\ref{eq:rho i limit}) and (\ref{eq:p i limit}) into (\ref{eq:Ricci scalar}) and (\ref{eq:omega 0}) to get
\begin{equation}
M\approx\frac{1}{a_{i}}\sqrt{\frac{4x-3}{4x^{2}-6x+3}}\,.
\label{eq:M limit}
\end{equation}
Here, we also made use of the fact that $\omega_{0}\ll 1$. Taking the $x\rightarrow\infty$ and $x\rightarrow 3/2$ limits, we obtain the range of scalar field masses for which the initial state undergoes accelerated expansion:
\begin{equation}
M\in\left(0,1/a_{i}\right)\,.
\label{eq:range of masses}
\end{equation}
This is perfectly consistent with a claim made by Hawking and Ellis at the end of section 4.3 in \cite{Hawking:1973uf}, where it is shown that the energy-momentum tensor of a scalar field minimally coupled to gravity globally satisfies the strong energy condition in a region of size $L$ if $L\gtrsim 1/M$. In other words, the convergence of timelike geodesics should be unaffected over distances greater than the Compton wavelength of the field. This is precisely what we have obtained: accelerated expansion (that is, violation of the strong energy condition) takes place if the radius of curvature, $a_{i}$, is smaller than $1/M$. 

\begin{figure}[h!]
\centering
\includegraphics[scale=0.605]{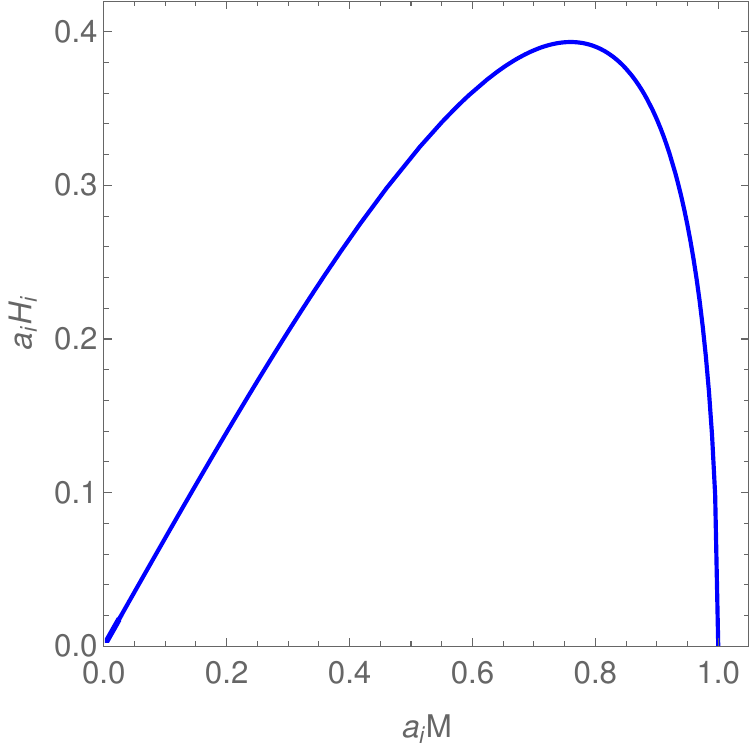}
\includegraphics[scale=0.62]{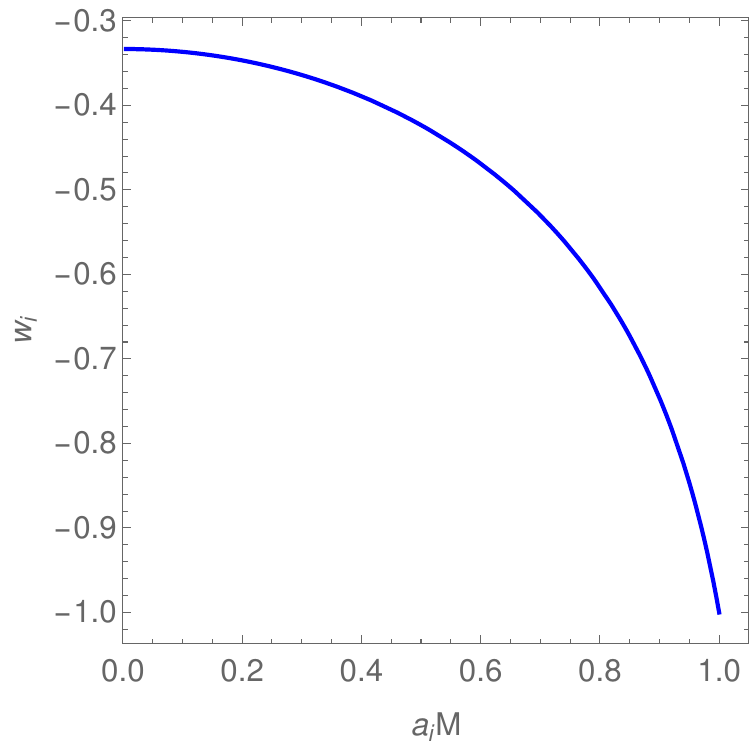}
\caption{Hubble rate (left) and equation of state parameter (right) corresponding to the different scalar field masses that allow for accelerated expansion.}
\label{fig:parametric}
\end{figure}

In order to better visualize our results, we provide in figure \ref{fig:parametric} parametric plots of (\ref{eq:w i limit}), (\ref{eq:H i limit}) and (\ref{eq:M limit}) showing what the induced Hubble rate and equation of state parameter are depending on the scalar field mass. It is interesting to note that there is a special mass $M_{*}$ for which the expansion rate is maximal. The value of $x$ that maximizes the positive solution in (\ref{eq:H i limit}) is $x_{*}=(3+\sqrt{3})/2$,
corresponding to
\begin{equation}
a_{i}M_{*}=\frac{1}{3^{1/4}}\approx 0.76\,.
\label{eq:M star}
\end{equation}
For this special mass, the Hubble rate and equation of state parameter are
\begin{equation}
H_{*}=\frac{M_{*}}{\sqrt{2+\sqrt{3}}}\approx 0.52M_{*}
\label{eq:H star}
\end{equation}
and
\begin{equation}
w_{*}=-\frac{1}{\sqrt{3}}\approx-0.58\,.
\label{eq:w star}
\end{equation}

\section{Conclusions}
\label{sec:conclusions}

In most models of inflation and dark energy, cosmic acceleration is driven by a homogeneous scalar field that evolves under the influence of a specially designed potential. Motivated by the search for a more natural mechanism, in this work we have proposed the possibility that accelerated expansion be caused instead by simple scalar field fluctuations about a minimum of the potential. In this picture, the universe is dominated by a gas of bosons forming a scalar field background radiation analogous to the CMB.

We first addressed this possibility in a spatially flat FLRW universe, showing that the well-motivated initial state given by (\ref{eq:expansion chi}) and (\ref{eq:expansion chi prime}) cannot induce accelerated expansion, regardless of the curvature coupling $\xi$ and the form of the scalar field spectrum $\langle|\alpha_{n}|^{2}\rangle$. We also showed how the usual $\langle\rho\rangle\propto T^{4}$ scaling is modified for thermal, non-conformal fields in the high-temperature limit. Even for minimal coupling, this law gets significant corrections for temperatures close to the Planck mass.

We then turned to the case of a spatially closed FLRW universe. We demonstrated that the initial state given by (\ref{eq:expansion psi closed}) and (\ref{eq:expansion psi prime closed}), naturally motivated by the equation of motion, is capable of inducing accelerated expansion. This effect is a direct consequence of the compact topology of the closed universe, as evidenced by the extra terms appearing in the expressions for the ensemble-averaged energy density and pressure relative to the flat case. Therefore, we may well interpret this topology-sensitive mechanism as a cosmological Casimir effect. 

Accelerated expansion can take place even for minimal coupling, provided the effective mass of the field is sufficiently small. In this case, reality of the Hubble rate requires that the Compton wavelength of the bosons (defined as the inverse of the bare mass) be greater than the scale factor. Under these conditions, the topological corrections make the pressure sufficiently negative, leading to repulsive gravity. The Hubble rate induced by such a state ranges from zero to roughly half the bare mass of the scalar field. Interestingly, when the Compton wavelength matches the curvature radius (that is, the scale factor), the Hubble rate vanishes and the equation of state parameter is -1, resembling a cosmic bounce or a quantum-driven onset of inflation.
  
Although our results are expressed in terms of an arbitrary scalar field spectrum $\langle|\alpha_{n}|^{2}\rangle$, special care must be taken when vacuum fluctuations are considered. In this case, the finite sums obtained after renormalization are not guaranteed to be positive and the conclusions would need to be reexamined.   
A careful treatment of renormalization effects is deferred to upcoming publications.
Future work will also be aimed at computing the time evolution of the initial states considered in this study and assessing the viability of the fluctuation-driven acceleration as a model to describe inflation and dark energy in our universe.


\section{Acknowledgments}

My work is supported in part by National Science 
Foundation grant 
PHY-2419848.
I am especially grateful to Larry Ford, Alan Guth and Diego Pavón for valuable insights and discussions. I also thank José Juan Blanco-Pillado, Georgios Fanaras, Jaume Garriga, Dražen Glavan, Mark Hertzberg, Tanvi Karwal and Ken Olum for helpful comments at various stages of this study.

\bibliography{bibliography_accelerated_expansion.bib}

@article{SupernovaSearchTeam:1998fmf,
    author = "Riess, Adam G. and others",
    collaboration = "Supernova Search Team",
    title = "{Observational evidence from supernovae for an accelerating universe and a cosmological constant}",
    eprint = "astro-ph/9805201",
    archivePrefix = "arXiv",
    doi = "10.1086/300499",
    journal = "Astron. J.",
    volume = "116",
    pages = "1009--1038",
    year = "1998"
}

@article{SupernovaCosmologyProject:1998vns,
    author = "Perlmutter, S. and others",
    collaboration = "Supernova Cosmology Project",
    title = "{Measurements of $\Omega$ and $\Lambda$ from 42 High Redshift Supernovae}",
    eprint = "astro-ph/9812133",
    archivePrefix = "arXiv",
    reportNumber = "LBNL-41801, LBL-41801",
    doi = "10.1086/307221",
    journal = "Astrophys. J.",
    volume = "517",
    pages = "565--586",
    year = "1999"
}

@article{RevModPhys.61.1,
  title = {The cosmological constant problem},
  author = {Weinberg, Steven},
  journal = {Rev. Mod. Phys.},
  volume = {61},
  issue = {1},
  pages = {1--23},
  numpages = {0},
  year = {1989},
  month = {Jan},
  publisher = {American Physical Society},
  doi = {10.1103/RevModPhys.61.1},
  url = {https://link.aps.org/doi/10.1103/RevModPhys.61.1}
}

@article{DESI:2025zgx,
    author = "Abdul Karim, M. and others",
    collaboration = "DESI",
    title = "{DESI DR2 results. II. Measurements of baryon acoustic oscillations and cosmological constraints}",
    eprint = "2503.14738",
    archivePrefix = "arXiv",
    primaryClass = "astro-ph.CO",
    reportNumber = "FERMILAB-PUB-25-0169-PPD",
    doi = "10.1103/tr6y-kpc6",
    journal = "Phys. Rev. D",
    volume = "112",
    number = "8",
    pages = "083515",
    year = "2025"
}

@article{DESI:2024mwx,
    author = "Adame, A. G. and others",
    collaboration = "DESI",
    title = "{DESI 2024 VI: cosmological constraints from the measurements of baryon acoustic oscillations}",
    eprint = "2404.03002",
    archivePrefix = "arXiv",
    primaryClass = "astro-ph.CO",
    reportNumber = "FERMILAB-PUB-24-0154-PPD",
    doi = "10.1088/1475-7516/2025/02/021",
    journal = "JCAP",
    volume = "02",
    pages = "021",
    year = "2025"
}

@article{Wang:2024vmw,
    author = "Wang, B. and Abdalla, E. and Atrio-Barandela, F. and Pav{\'o}n, D.",
    title = "{Further understanding the interaction between dark energy and dark matter: current status and future directions}",
    eprint = "2402.00819",
    archivePrefix = "arXiv",
    primaryClass = "astro-ph.CO",
    doi = "10.1088/1361-6633/ad2527",
    journal = "Rept. Prog. Phys.",
    volume = "87",
    number = "3",
    pages = "036901",
    year = "2024"
}

@article{Li:2024qso,
    author = "Li, Tian-Nuo and Wu, Peng-Ju and Du, Guo-Hong and Jin, Shang-Jie and Li, Hai-Li and Zhang, Jing-Fei and Zhang, Xin",
    title = "{Constraints on Interacting Dark Energy Models from the DESI Baryon Acoustic Oscillation and DES Supernovae Data}",
    eprint = "2407.14934",
    archivePrefix = "arXiv",
    primaryClass = "astro-ph.CO",
    doi = "10.3847/1538-4357/ad87f0",
    journal = "Astrophys. J.",
    volume = "976",
    number = "1",
    pages = "1",
    year = "2024"
}

@article{Chakraborty:2025syu,
    author = "Chakraborty, Amlan and Chanda, Prolay K. and Das, Subinoy and Dutta, Koushik",
    title = "{DESI results: hint towards coupled dark matter and dark energy}",
    eprint = "2503.10806",
    archivePrefix = "arXiv",
    primaryClass = "astro-ph.CO",
    reportNumber = "JCAP11(2025)047",
    doi = "10.1088/1475-7516/2025/11/047",
    journal = "JCAP",
    volume = "11",
    pages = "047",
    year = "2025"
}

@article{Khoury:2025txd,
    author = "Khoury, Justin and Lin, Meng-Xiang and Trodden, Mark",
    title = "{Apparent w{\ensuremath{<}}-1 and a Lower S8 from Dark Axion and Dark Baryons Interactions}",
    eprint = "2503.16415",
    archivePrefix = "arXiv",
    primaryClass = "astro-ph.CO",
    doi = "10.1103/w4qb-plk8",
    journal = "Phys. Rev. Lett.",
    volume = "135",
    number = "18",
    pages = "181001",
    year = "2025"
}

@article{Guedezounme:2025wav,
    author = "Guedezounme, S{\^e}cloka L. and Dinda, Bikash R. and Maartens, Roy",
    title = "{Phantom crossing or dark interaction?}",
    eprint = "2507.18274",
    archivePrefix = "arXiv",
    primaryClass = "astro-ph.CO",
    doi = "10.1088/1475-7516/2026/01/062",
    journal = "JCAP",
    volume = "01",
    pages = "062",
    year = "2026"
}

@article{Wolf:2025jed,
    author = "Wolf, William J. and Garc{\'\i}a-Garc{\'\i}a, Carlos and Anton, Theodore and Ferreira, Pedro G.",
    title = "{Assessing Cosmological Evidence for Nonminimal Coupling}",
    eprint = "2504.07679",
    archivePrefix = "arXiv",
    primaryClass = "astro-ph.CO",
    doi = "10.1103/jysf-k72m",
    journal = "Phys. Rev. Lett.",
    volume = "135",
    number = "8",
    pages = "081001",
    year = "2025"
}

@article{Bayat:2025xfr,
    author = "Bayat, Zahra and Hertzberg, Mark P.",
    title = "{Examining quintessence models with DESI data}",
    eprint = "2505.18937",
    archivePrefix = "arXiv",
    primaryClass = "astro-ph.CO",
    doi = "10.1088/1475-7516/2025/08/065",
    journal = "JCAP",
    volume = "08",
    pages = "065",
    year = "2025"
}

@article{Gialamas:2025pwv,
    author = {Gialamas, Ioannis D. and H{\"u}tsi, Gert and Raidal, Martti and Urrutia, Juan and Vasar, Martin and Veerm{\"a}e, Hardi},
    title = "{Quintessence and phantoms in light of DESI 2025}",
    eprint = "2506.21542",
    archivePrefix = "arXiv",
    primaryClass = "astro-ph.CO",
    doi = "10.1103/kdqc-y37v",
    journal = "Phys. Rev. D",
    volume = "112",
    number = "6",
    pages = "063551",
    year = "2025"
}

@article{Wolf:2024eph,
    author = "Wolf, William J. and Garc{\'\i}a-Garc{\'\i}a, Carlos and Bartlett, Deaglan J. and Ferreira, Pedro G.",
    title = "{Scant evidence for thawing quintessence}",
    eprint = "2408.17318",
    archivePrefix = "arXiv",
    primaryClass = "astro-ph.CO",
    doi = "10.1103/PhysRevD.110.083528",
    journal = "Phys. Rev. D",
    volume = "110",
    number = "8",
    pages = "083528",
    year = "2024"
}

@article{Wolf:2024stt,
    author = "Wolf, William J. and Ferreira, Pedro G. and Garc{\'\i}a-Garc{\'\i}a, Carlos",
    title = "{Matching current observational constraints with nonminimally coupled dark energy}",
    eprint = "2409.17019",
    archivePrefix = "arXiv",
    primaryClass = "astro-ph.CO",
    doi = "10.1103/PhysRevD.111.L041303",
    journal = "Phys. Rev. D",
    volume = "111",
    number = "4",
    pages = "L041303",
    year = "2025"
}

@article{Ye:2024ywg,
    author = "Ye, Gen and Martinelli, Matteo and Hu, Bin and Silvestri, Alessandra",
    title = "{Hints of Nonminimally Coupled Gravity in DESI 2024 Baryon Acoustic Oscillation Measurements}",
    eprint = "2407.15832",
    archivePrefix = "arXiv",
    primaryClass = "astro-ph.CO",
    doi = "10.1103/PhysRevLett.134.181002",
    journal = "Phys. Rev. Lett.",
    volume = "134",
    number = "18",
    pages = "181002",
    year = "2025"
}

@article{Adam:2025kve,
    author = "Adam, Husam and Hertzberg, Mark P. and Jim{\'e}nez-Aguilar, Daniel and Khan, Iman",
    title = "{Comparing minimal and non-minimal quintessence models to 2025 DESI data}",
    eprint = "2509.13302",
    archivePrefix = "arXiv",
    primaryClass = "astro-ph.CO",
    doi = "10.1088/1475-7516/2026/04/052",
    journal = "JCAP",
    volume = "04",
    pages = "052",
    year = "2026"
}

@article{Planck:2018jri,
    author = "Akrami, Y. and others",
    collaboration = "Planck",
    title = "{Planck 2018 results. X. Constraints on inflation}",
    eprint = "1807.06211",
    archivePrefix = "arXiv",
    primaryClass = "astro-ph.CO",
    doi = "10.1051/0004-6361/201833887",
    journal = "Astron. Astrophys.",
    volume = "641",
    pages = "A10",
    year = "2020"
}

@article{Guth:1980zm,
    author = "Guth, Alan H.",
    editor = "Fang, Li-Zhi and Ruffini, R.",
    title = "{The Inflationary Universe: A Possible Solution to the Horizon and Flatness Problems}",
    reportNumber = "SLAC-PUB-2576",
    doi = "10.1103/PhysRevD.23.347",
    journal = "Phys. Rev. D",
    volume = "23",
    pages = "347--356",
    year = "1981"
}

@article{Linde:1981mu,
    author = "Linde, Andrei D.",
    editor = "Fang, Li-Zhi and Ruffini, R.",
    title = "{A New Inflationary Universe Scenario: A Possible Solution of the Horizon, Flatness, Homogeneity, Isotropy and Primordial Monopole Problems}",
    reportNumber = "LEBEDEV-81-229",
    doi = "10.1016/0370-2693(82)91219-9",
    journal = "Phys. Lett. B",
    volume = "108",
    pages = "389--393",
    year = "1982"
}

@article{Albrecht:1982wi,
    author = "Albrecht, Andreas and Steinhardt, Paul J.",
    editor = "Fang, Li-Zhi and Ruffini, R.",
    title = "{Cosmology for Grand Unified Theories with Radiatively Induced Symmetry Breaking}",
    reportNumber = "UPR-0185T",
    doi = "10.1103/PhysRevLett.48.1220",
    journal = "Phys. Rev. Lett.",
    volume = "48",
    pages = "1220--1223",
    year = "1982"
}

@article{SDSS:2003eyi,
    author = "Tegmark, Max and others",
    collaboration = "SDSS",
    title = "{Cosmological parameters from SDSS and WMAP}",
    eprint = "astro-ph/0310723",
    archivePrefix = "arXiv",
    reportNumber = "FERMILAB-PUB-03-435-A",
    doi = "10.1103/PhysRevD.69.103501",
    journal = "Phys. Rev. D",
    volume = "69",
    pages = "103501",
    year = "2004"
}

@article{Guth:2005zr,
    author = "Guth, Alan H. and Kaiser, David I.",
    title = "{Inflationary cosmology: Exploring the Universe from the smallest to the largest scales}",
    eprint = "astro-ph/0502328",
    archivePrefix = "arXiv",
    reportNumber = "MIT-CTP-3594",
    doi = "10.1126/science.1107483",
    journal = "Science",
    volume = "307",
    pages = "884--890",
    year = "2005"
}

@book{Hawking:1973uf,
    author = "Hawking, Stephen W. and Ellis, George F. R.",
    title = "{The Large Scale Structure of Space-Time}",
    doi = "10.1017/9781009253161",
    isbn = "978-1-009-25316-1, 978-1-009-25315-4, 978-0-521-20016-5, 978-0-521-09906-6, 978-0-511-82630-6, 978-0-521-09906-6",
    publisher = "Cambridge University Press",
    series = "Cambridge Monographs on Mathematical Physics",
    month = "2",
    year = "2023"
}

@book{Birrell:1982ix,
    author = "Birrell, N. D. and Davies, P. C. W.",
    title = "{Quantum Fields in Curved Space}",
    doi = "10.1017/CBO9780511622632",
    isbn = "978-0-511-62263-2, 978-0-521-27858-4",
    publisher = "Cambridge University Press",
    address = "Cambridge, UK",
    series = "Cambridge Monographs on Mathematical Physics",
    year = "1982"
}

@book{Mukhanov:2007zz,
    author = "Mukhanov, Viatcheslav and Winitzki, Sergei",
    title = "{Introduction to quantum effects in gravity}",
    isbn = "978-0-521-86834-1, 978-1-139-78594-5",
    publisher = "Cambridge University Press",
    month = "6",
    year = "2007"
}

@article{PhysRevD.9.341,
  title = {Adiabatic regularization of the energy-momentum tensor of a quantized field in homogeneous spaces},
  author = {Parker, Leonard and Fulling, S. A.},
  journal = {Phys. Rev. D},
  volume = {9},
  issue = {2},
  pages = {341--354},
  numpages = {0},
  year = {1974},
  month = {Jan},
  publisher = {American Physical Society},
  doi = {10.1103/PhysRevD.9.341},
  url = {https://link.aps.org/doi/10.1103/PhysRevD.9.341}
}

@article{Ford:1976fn,
    author = "Ford, L. H.",
    title = "{Quantum Vacuum Energy in a Closed Universe}",
    doi = "10.1103/PhysRevD.14.3304",
    journal = "Phys. Rev. D",
    volume = "14",
    pages = "3304--3313",
    year = "1976"
}

@article{Ford:1975su,
    author = "Ford, L. H.",
    title = "{Quantum Vacuum Energy in General Relativity}",
    doi = "10.1103/PhysRevD.11.3370",
    journal = "Phys. Rev. D",
    volume = "11",
    pages = "3370--3377",
    year = "1975"
}

@article{Streeruwitz:1975wzf,
    author = "Streeruwitz, E.",
    title = "{Vacuum fluctuations of a scalar field in an Einstein universe}",
    doi = "10.1016/0370-2693(75)90195-1",
    journal = "Phys. Lett. B",
    volume = "55",
    pages = "93--96",
    year = "1975"
}

@article{Streeruwitz:1975er,
    author = "Streeruwitz, E.",
    title = "{Vacuum Fluctuations of a Quantized Scalar Field in a Robertson-Walker Universe}",
    doi = "10.1016/0370-2693(75)90500-6",
    journal = "Phys. Lett. B",
    volume = "56",
    pages = "66--68",
    year = "1975"
}

@article{Dowker:1976pr,
    author = "Dowker, J. S. and Critchley, Raymond",
    title = "{Vacuum Stress Tensor in an Einstein Universe. Finite Temperature Effects}",
    reportNumber = "Print-76-0588 (MANCHESTER)",
    doi = "10.1103/PhysRevD.15.1484",
    journal = "Phys. Rev. D",
    volume = "15",
    pages = "1484",
    year = "1977"
}

@article{Dowker:1978fr,
    author = "Dowker, J. S. and Altaie, B. M.",
    title = "{Spinor Fields in an Einstein Universe. The Vacuum Averaged Stress Energy Tensor}",
    doi = "10.1103/PhysRevD.17.417",
    journal = "Phys. Rev. D",
    volume = "17",
    pages = "417--422",
    year = "1978"
}

@article{Russell:1987qm,
    author = "Russell, I. H. and Toms, D. J.",
    title = "{Vacuum Energy for Massive Forms in R(m) X S($N$)}",
    reportNumber = "NCL-87-TP5",
    doi = "10.1088/0264-9381/4/5/030",
    journal = "Class. Quant. Grav.",
    volume = "4",
    pages = "1357",
    year = "1987"
}

@article{Herdeiro:2005zj,
    author = "Herdeiro, Carlos A. R. and Sampaio, Marco",
    title = "{Casimir energy and a cosmological bounce}",
    eprint = "hep-th/0510052",
    archivePrefix = "arXiv",
    doi = "10.1088/0264-9381/23/2/012",
    journal = "Class. Quant. Grav.",
    volume = "23",
    pages = "473--484",
    year = "2006"
}

@article{Herdeiro:2009zza,
    author = "Herdeiro, Carlos A. R. and Ribeiro, Raquel H. and Sampaio, Marco",
    editor = "Bezerra, V. B. and Mostepanenko, V. M. and Romero, Carlos",
    title = "{Repulsive gravity and the Casimir effect on spherical universes}",
    doi = "10.1142/S0217751X09045406",
    journal = "Int. J. Mod. Phys. A",
    volume = "24",
    pages = "1821--1824",
    year = "2009"
}

@article{Saharian:2010nep,
    author = "Saharian, A. A. and Mkhitaryan, A. L.",
    title = "{Vacuum fluctuations and topological Casimir effect in Friedmann-Robertson-Walker cosmologies with compact dimensions}",
    eprint = "0908.3291",
    archivePrefix = "arXiv",
    primaryClass = "hep-th",
    doi = "10.1140/epjc/s10052-010-1247-0",
    journal = "Eur. Phys. J. C",
    volume = "66",
    pages = "295--306",
    year = "2010"
}

@article{Elizalde:2003ke,
    author = "Elizalde, E. and Tort, A. C.",
    title = "{A Note on the Casimir energy of a massive scalar field in positive curvature space}",
    eprint = "hep-th/0306049",
    archivePrefix = "arXiv",
    doi = "10.1142/S0217732304012836",
    journal = "Mod. Phys. Lett. A",
    volume = "19",
    pages = "111--116",
    year = "2004"
}

@article{Bordag:2001qi,
    author = "Bordag, Michael and Mohideen, U. and Mostepanenko, V. M.",
    title = "{New developments in the Casimir effect}",
    eprint = "quant-ph/0106045",
    archivePrefix = "arXiv",
    doi = "10.1016/S0370-1573(01)00015-1",
    journal = "Phys. Rept.",
    volume = "353",
    pages = "1--205",
    year = "2001"
}

@article{Mamaev:1976zb,
    author = "Mamaev, S. G. and Mostepanenko, V. M. and Starobinsky, Alexei A.",
    title = "{Particle creation from the vacuum near a homogeneous isotropic singularity}",
    journal = "Sov. Phys. JETP",
    volume = "43",
    pages = "823--830",
    year = "1976"
}

@book{Mostepanenko:1990ceg,
    author = "Mostepanenko, V. M. and Trunov, N. N.",
    title = "{The Casimir effect and its applications}",
    isbn = "978-0-19-853998-8, 978-0-19-853998-8, 978-5-283-03922-0",
    publisher = "Clarendon Press",
    address = "Oxford, New York",
    year = "1997"
}

@book{Krech1994,
  author    = {Michael Krech},
  title     = {The Casimir Effect in Critical Systems},
  publisher = {World Scientific},
  address   = {Singapore},
  year      = {1994},
  isbn      = {9789810218454}
}

@article{Casimir:1948dh,
    author = "Casimir, H. B. G.",
    title = "{On the attraction between two perfectly conducting plates}",
    journal = "Indag. Math.",
    volume = "10",
    number = "4",
    pages = "261--263",
    year = "1948"
}

@article{Zeldovich:1984vk,
    author = "Zeldovich, Ya. B. and Starobinsky, Alexei A.",
    title = "{Quantum creation of a universe in a nontrivial topology}",
    journal = "Sov. Astron. Lett.",
    volume = "10",
    pages = "135",
    year = "1984"
}

@article{Bezerra:2011nc,
    author = "Bezerra, V. B. and Mostepanenko, V. M. and Mota, H. F. and Romero, C.",
    title = "{Thermal Casimir effect for neutrino and electromagnetic fields in closed Friedmann cosmological model}",
    eprint = "1110.4504",
    archivePrefix = "arXiv",
    primaryClass = "gr-qc",
    doi = "10.1103/PhysRevD.84.104025",
    journal = "Phys. Rev. D",
    volume = "84",
    pages = "104025",
    year = "2011"
}

@article{PhysRevD.83.104042,
  title = {Thermal Casimir effect in closed Friedmann universe revisited},
  author = {Bezerra, V. B. and Klimchitskaya, G. L. and Mostepanenko, V. M. and Romero, C.},
  journal = {Phys. Rev. D},
  volume = {83},
  issue = {10},
  pages = {104042},
  numpages = {7},
  year = {2011},
  month = {May},
  publisher = {American Physical Society},
  doi = {10.1103/PhysRevD.83.104042},
  url = {https://link.aps.org/doi/10.1103/PhysRevD.83.104042}
}

@article{Zhuk:1996xc,
    author = "Zhuk, A. and Kleinert, H.",
    title = "{Casimir effect at nonzero temperatures in a closed Friedmann universe}",
    doi = "10.1007/BF02072013",
    journal = "Theor. Math. Phys.",
    volume = "109",
    pages = "1483--1493",
    year = "1996"
}

@article{Bezerra:2024wwm,
    author = "Bezerra, Valdir Barbosa and Santana Mota, Herondy Francisco and Lima, Augusto P. C. M. and Alencar, Geov{\'a} and Muniz, Celio Rodrigues",
    title = "{The Casimir Effect in Finite-Temperature and Gravitational Scenarios}",
    doi = "10.3390/physics6030065",
    journal = "MDPI Physics",
    volume = "6",
    number = "3",
    pages = "1046--1071",
    year = "2024"
}

@article{Bezerra:2017zqq,
    author = "Bezerra, V. B. and Mota, H. F. and Muniz, C. R.",
    title = "{Casimir Effect in the Rainbow Einstein's Universe}",
    eprint = "1708.02627",
    archivePrefix = "arXiv",
    primaryClass = "gr-qc",
    doi = "10.1209/0295-5075/120/10005",
    journal = "EPL",
    volume = "120",
    number = "1",
    pages = "10005",
    year = "2017"
}

@article{Cho:2022ngf,
    author = "Cho, Hing-Tong and Hsiang, Jen-Tsung and Hu, Bei-Lok",
    title = "{Quantum Capacity and Vacuum Compressibility of Spacetime: Thermal Fields}",
    eprint = "2204.08634",
    archivePrefix = "arXiv",
    primaryClass = "gr-qc",
    doi = "10.3390/universe8050291",
    journal = "Universe",
    volume = "8",
    number = "5",
    pages = "291",
    year = "2022"
}

@article{Xie:2023yfg,
    author = "Xie, Yu-Cun and Hsiang, Jen-Tsung and Hu, Bei-Lok",
    title = "{Dynamical vacuum compressibility of space}",
    eprint = "2312.09047",
    archivePrefix = "arXiv",
    primaryClass = "gr-qc",
    doi = "10.1103/PhysRevD.109.065027",
    journal = "Phys. Rev. D",
    volume = "109",
    number = "6",
    pages = "065027",
    year = "2024"
}

@article{Hsiang:2024keo,
    author = "Hsiang, Jen-Tsung and Xie, Yu-Cun and Hu, Bei-Lok",
    title = "{Heat capacity and quantum compressibility of dynamical spacetimes with thermal particle creation}",
    eprint = "2405.00360",
    archivePrefix = "arXiv",
    primaryClass = "hep-th",
    doi = "10.1103/PhysRevD.110.063504",
    journal = "Phys. Rev. D",
    volume = "110",
    number = "6",
    pages = "063504",
    year = "2024"
}

@article{Szydlowski:2007bg,
    author = "Szydlowski, Marek and Godlowski, Wlodzimierz",
    title = "{Acceleration of the universe driven by the Casimir force}",
    eprint = "0705.1772",
    archivePrefix = "arXiv",
    primaryClass = "gr-qc",
    doi = "10.1142/S021827180801205X",
    journal = "Int. J. Mod. Phys. D",
    volume = "17",
    pages = "343--366",
    year = "2008"
}

@article{Moreira:2025cwp,
    author = "Moreira, Jr., E. S. and Paula, J. P. A.",
    title = "{Thermodynamic stability in an Einstein universe}",
    eprint = "2512.23838",
    archivePrefix = "arXiv",
    primaryClass = "gr-qc",
    doi = "10.1103/zcml-wwcz",
    journal = "Phys. Rev. D",
    volume = "113",
    number = "10",
    pages = "105004",
    year = "2026"
}

@article{Hertzberg:2025ifp,
    author = "Hertzberg, Mark P. and Jim{\'e}nez-Aguilar, Daniel",
    title = "{Probability of the initial conditions for inflation and slow contraction}",
    eprint = "2505.22763",
    archivePrefix = "arXiv",
    primaryClass = "gr-qc",
    doi = "10.1103/klnz-mjj2",
    journal = "Phys. Rev. D",
    volume = "112",
    number = "4",
    pages = "043519",
    year = "2025"
}

@article{Steinhardt:2011zza,
    author = "Steinhardt, Paul J.",
    title = "{The inflation debate: Is the theory at the heart of modern cosmology deeply flawed?}",
    journal = "Sci. Am.",
    volume = "304N4",
    pages = "18--25",
    year = "2011"
}

@article{Ijjas:2024oqn,
    author = "Ijjas, Anna and Steinhardt, Paul J. and Garfinkle, David and Cook, William G.",
    title = "{Smoothing and flattening the universe through slow contraction versus inflation}",
    eprint = "2404.00867",
    archivePrefix = "arXiv",
    primaryClass = "gr-qc",
    doi = "10.1088/1475-7516/2024/07/077",
    journal = "JCAP",
    volume = "07",
    pages = "077",
    year = "2024"
}

@article{Aurrekoetxea:2024ypv,
    author = "Aurrekoetxea, Josu C. and Clough, Katy and Lim, Eugene A.",
    title = "{Cosmology using numerical relativity}",
    eprint = "2409.01939",
    archivePrefix = "arXiv",
    primaryClass = "gr-qc",
    doi = "10.1007/s41114-025-00058-z",
    journal = "Living Rev. Rel.",
    volume = "28",
    number = "1",
    pages = "5",
    year = "2025"
}

@article{Hartle:1983ai,
    author = "Hartle, J. B. and Hawking, S. W.",
    editor = "Fang, Li-Zhi and Ruffini, R.",
    title = "{Wave Function of the Universe}",
    reportNumber = "PRINT-83-0937 (CAMBRIDGE)",
    doi = "10.1103/PhysRevD.28.2960",
    journal = "Phys. Rev. D",
    volume = "28",
    pages = "2960--2975",
    year = "1983"
}

@article{Vilenkin:1984wp,
    author = "Vilenkin, A.",
    title = "{Quantum Creation of Universes}",
    doi = "10.1103/PhysRevD.30.509",
    journal = "Phys. Rev. D",
    volume = "30",
    pages = "509--511",
    year = "1984"
}

@article{Mamaev:1979ks,
    author = "Mamaev, S. G. and Trunov, N. N.",
    title = "{DEPENDENCE OF THE VACUUM EXPECTATION VALUES OF THE ENERGY MOMENTUM TENSOR ON THE GEOMETRY AND TOPOLOGY OF THE MANIFOLD}",
    doi = "10.1007/BF01018540",
    journal = "Theor. Math. Phys.",
    volume = "38",
    pages = "228--234",
    year = "1979"
}

@article{Mamaev:1979zw,
    author = "Mamaev, S. G. and Trunov, N. N.",
    title = "{VACUUM AVERAGES OF THE ENERGY MOMENTUM TENSOR OF QUANTIZED FIELDS ON MANIFOLDS OF VARIOUS TOPOLOGY AND GEOMETRY. II}",
    doi = "10.1007/BF00891392",
    journal = "Sov. Phys. J.",
    volume = "22",
    pages = "966--968",
    year = "1979"
}

@article{BezerradeMello:2017nyo,
    author = "Bezerra de Mello, E. R. and Saharian, A. A. and Setare, M. R.",
    title = "{The Casimir effect for parallel plates in Friedmann-Robertson-Walker universe}",
    eprint = "1701.05026",
    archivePrefix = "arXiv",
    primaryClass = "hep-th",
    doi = "10.1103/PhysRevD.95.065024",
    journal = "Phys. Rev. D",
    volume = "95",
    number = "6",
    pages = "065024",
    year = "2017"
}

@inproceedings{Guth:2025dal,
    author = "Guth, Alan H. and Vilenkin, Alexander",
    title = "{On quantum creation of a toroidal universe}",
    eprint = "2508.08747",
    archivePrefix = "arXiv",
    primaryClass = "gr-qc",
    reportNumber = "MIT-CTP-LI/5883",
    journal = "",
    month = "8",
    year = "2025"
}

@article{SanchezLopez:2025uzw,
    author = "S{\'a}nchez L{\'o}pez, Samuel and Karam, Alexandros and Hazra, Dhiraj Kumar",
    title = "{Non-Minimally Coupled Quintessence in Light of DESI}",
    eprint = "2510.14941",
    archivePrefix = "arXiv",
    primaryClass = "astro-ph.CO",
    journal = "",
    month = "10",
    year = "2025"
}

@article{Li:2026xaz,
    author = "Li, Tian-Nuo and Giar{\`e}, William and Du, Guo-Hong and Li, Yun-He and Di Valentino, Eleonora and Zhang, Jing-Fei and Zhang, Xin",
    title = "{Strong Evidence for Dark Sector Interactions}",
    eprint = "2601.07361",
    archivePrefix = "arXiv",
    primaryClass = "astro-ph.CO",
    journal = "",
    month = "1",
    year = "2026"
}

@article{Shlivko:2024llw,
    author = "Shlivko, David and Steinhardt, Paul J.",
    title = "{Assessing observational constraints on dark energy}",
    eprint = "2405.03933",
    archivePrefix = "arXiv",
    primaryClass = "astro-ph.CO",
    doi = "10.1016/j.physletb.2024.138826",
    journal = "Phys. Lett. B",
    volume = "855",
    pages = "138826",
    year = "2024"
}

@article{Akrami:2025zlb,
    author = "Akrami, Yashar and Alestas, George and Nesseris, Savvas",
    title = "{Has DESI detected exponential quintessence?}",
    eprint = "2504.04226",
    archivePrefix = "arXiv",
    primaryClass = "astro-ph.CO",
    reportNumber = "IFT-UAM/CSIC-25-36",
    journal = "",
    month = "4",
    year = "2025"
}

@article{Cline:2025sbt,
    author = "Cline, James M. and Muralidharan, Varun",
    title = "{Simple quintessence models in light of DESI-BAO observations}",
    eprint = "2506.13047",
    archivePrefix = "arXiv",
    primaryClass = "astro-ph.CO",
    doi = "10.1103/8z2m-nbv6",
    journal = "Phys. Rev. D",
    volume = "112",
    number = "6",
    pages = "063539",
    year = "2025"
}

@article{LaPenna:2026avs,
    author = "La Penna, Lorenzo and Notari, Alessio and Redi, Michele",
    title = "{Mimicking Phantom Dark Energy with Evolving Dark Matter Mass}",
    eprint = "2601.05235",
    archivePrefix = "arXiv",
    primaryClass = "astro-ph.CO",
    journal = "",
    month = "1",
    year = "2026"
}

\appendix

\section{Computation of ensemble averages}
\label{sec:appendix ensemble averages}
In this appendix, we will give the necessary details to derive the ensemble averages (\ref{eq:mean 1})-(\ref{eq:mean 4}), corresponding to a spatially flat universe. In the case of a closed universe, the Laplacian has a discrete spectrum labeled by $n$, with eigenvalues $n(n+2)$ and degeneracy $(n+1)^{2}$. Therefore, one can obtain the closed universe averages (\ref{eq:mean 1 closed})-(\ref{eq:mean 4 closed}) from the flat case by making these two replacements:
\begin{equation}
\vec{q}^{\,\,2}\rightarrow n(n+2)\,,
\end{equation}
\begin{equation}
\frac{1}{(2\pi)^{3}}\int d^{3}q\rightarrow\frac{1}{2\pi^{2}}\sum_{n=0}^{\infty}(n+1)^{2}\,.
\end{equation}
\\
Let $f_{\psi}(\vec{q})$ be the Fourier transform of $\psi(\eta_i,\vec{x})$, namely,
\begin{equation}
\psi(\eta_i,\vec{x})=\int d^{3}q f_{\psi}(\vec{q})e^{i\vec{q}\cdot\vec{x}}\,.
\label{eq:Fourier psi}
\end{equation}
Similarly, $f_{\pi}(\vec{q})$ will denote the Fourier transform of $\psi'(\eta_i,\vec{x})$:
\begin{equation}
\psi'(\eta_i,\vec{x})=\int d^{3}q f_{\pi}(\vec{q})e^{i\vec{q}\cdot\vec{x}}\,.
\end{equation}
Comparison of these two expressions with (\ref{eq:expansion chi}) and (\ref{eq:expansion chi prime}) yields
\begin{equation}
f_{\psi}(\vec{q})=\frac{1}{\sqrt{2(2\pi)^{3}\omega_{\vec{q}}}}\left(\alpha_{\vec{q}}+\alpha_{\vec{q}}^{*}\right)\,,
\end{equation}
\begin{equation}
f_{\pi}(\vec{q})=-i\sqrt{\frac{\omega_{\vec{q}}}{2(2\pi)^{3}}}\left(\alpha_{\vec{q}}-\alpha_{\vec{q}}^{*}\right)\,.
\end{equation}
As pointed out in the main text, the real and imaginary parts of $\alpha_{\vec{q}}$ ($A_{\vec{q}}$ and $B_{\vec{q}}$, respectively) are independent random variables with vanishing expectation value and variance $\langle|\alpha_{\vec{q}}|^{2}\rangle/2$, meaning that
\begin{equation}
\langle A_{\vec{q}}\rangle=\langle B_{\vec{p}}\rangle=0\,,
\end{equation}
\begin{equation}
\langle A_{\vec{q}}B_{\vec{p}}\rangle=0\,,
\end{equation}
\begin{equation}
\langle A_{\vec{q}}A_{\vec{p}}\rangle=\langle B_{\vec{q}}B_{\vec{p}}\rangle=\frac{\langle|\alpha_{\vec{q}}|^{2}\rangle}{2}\delta^{(3)}(\vec{q}-\vec{p})\,.
\end{equation}
One can use these expressions to find $\langle f_{\psi}(\vec{q})f_{\psi}(\vec{p})\rangle$, $\langle f_{\pi}(\vec{q})f_{\pi}(\vec{p})\rangle$ and $\langle f_{\psi}(\vec{q})f_{\pi}(\vec{p})\rangle$. After some straightforward algebra, we get
\begin{equation}
\langle f_{\psi}(\vec{q})f_{\psi}(\vec{p})\rangle=\frac{\langle|\alpha_{\vec{q}}|^{2}\rangle}{(2\pi)^{3}\sqrt{\omega_{\vec{q}}\omega_{\vec{p}}}}\delta^{(3)}(\vec{q}+\vec{p})\,,
\label{eq:power fpsi}
\end{equation}
\begin{equation}
\langle f_{\pi}(\vec{q})f_{\pi}(\vec{p})\rangle=\frac{\sqrt{\omega_{\vec{q}}\omega_{\vec{p}}}}{(2\pi)^{3}}\langle|\alpha_{\vec{q}}|^{2}\rangle\delta^{(3)}(\vec{q}+\vec{p})\,,
\label{eq:power fpi}
\end{equation}
\begin{equation}
\langle f_{\psi}(\vec{q})f_{\pi}(\vec{p})\rangle=0\,.
\label{eq:power fpsi fpi}
\end{equation}
With this information, we can now compute the averages (\ref{eq:mean 1})-(\ref{eq:mean 4}):
\begin{itemize}
\item $\langle\psi^{2}\rangle$: \\\\
According to the convolution theorem, the Fourier transform of $\psi^{2}$ is
\begin{equation}
f_{\psi^{2}}(\vec{p})=\int d^{3}q f_{\psi}(\vec{q})f_{\psi}(\vec{p}-\vec{q})\,.
\end{equation}
Therefore,
\begin{equation}
\langle\psi^{2}\rangle=\int d^{3}p\int d^{3}q\,\langle f_{\psi}(\vec{q})f_{\psi}(\vec{p}-\vec{q})\rangle e^{i\vec{p}\cdot\vec{x}}\,.
\end{equation}
Using (\ref{eq:power fpsi}) and computing the $d^{3}p$ integral, we get
\begin{equation}
\langle\psi^{2}\rangle=\frac{1}{(2\pi)^3}\int d^{3}q\,\frac{\langle|\alpha_{\vec{q}}|^{2}\rangle}{\omega_{\vec{q}}}\,.
\end{equation}
\item $\langle(\psi')^{2}\rangle$:\\\\
Applying exactly the same procedure with $f_{\pi}$ instead of $f_{\psi}$, we get
\begin{equation}
\langle(\psi')^{2}\rangle=\frac{1}{(2\pi)^3}\int d^{3}q\,\omega_{\vec{q}}\langle|\alpha_{\vec{q}}|^{2}\rangle\,.
\end{equation}
\item $\langle(\vec{\nabla}\psi)^2\rangle$:\\\\
The square of the gradient of $\psi$ is given by
\begin{equation}
(\vec{\nabla}\psi)^2=a^{2}g^{ij}\partial_{i}\psi\partial_{j}\psi=(\partial_{\chi}\psi)^{2}+\frac{(\partial_{\theta}\psi)^{2}}{\Sigma^{2}(\chi)}+\frac{(\partial_{\varphi}\psi)^{2}}{\Sigma^{2}(\chi)\sin^{2}\theta}\,.
\label{eq:gradient squared}
\end{equation}
The Fourier transform of $\partial_{i}\psi$ can be easily deduced by taking the derivative of (\ref{eq:Fourier psi}) with respect to $x^{i}$:
\begin{equation}
f_{\psi_{,i}}(\vec{p})=ip_{i}f_{\psi}(\vec{p})\,.
\label{eq:f psi comma i}
\end{equation}
Now, by the convolution theorem, the Fourier transform of $(\partial_{i}\psi)^{2}$ is
\begin{equation}
f_{\psi_{,i}^{2}}(\vec{p})=\int d^{3}q f_{\psi_{,i}}(\vec{q})f_{\psi_{,i}}(\vec{p}-\vec{q})=\int d^{3}q\,q_{i}(q_{i}-p_{i})f_{\psi}(\vec{q})f_{\psi}(\vec{p}-\vec{q})\,,
\end{equation}
so
\begin{equation}
\langle(\partial_{i}\psi)^{2}\rangle=\int d^{3}p\int d^{3}q\,q_{i}(q_{i}-p_{i})\langle f_{\psi}(\vec{q})f_{\psi}(\vec{p}-\vec{q})\rangle e^{i\vec{p}\cdot\vec{x}}\,.
\end{equation}
Using (\ref{eq:power fpsi}) and computing the $d^{3}p$ integral, we get
\begin{equation}
\langle(\partial_{i}\psi)^{2}\rangle=\frac{1}{(2\pi)^3}\int d^{3}q\,\frac{q_{i}^{2}\langle|\alpha_{\vec{q}}|^{2}\rangle}{\omega_{\vec{q}}}\,.
\end{equation}
Finally, inserting this into (\ref{eq:gradient squared}) and taking into account that $\vec{q}^{\,\,2}=a^{2}g^{ij}q_{i}q_{j}$, we get
\begin{equation}
\langle(\vec{\nabla}\psi)^2\rangle=\frac{1}{(2\pi)^3}\int d^{3}q\,\frac{\vec{q}^{\,\,2}\langle|\alpha_{\vec{q}}|^{2}\rangle}{\omega_{\vec{q}}}\,.
\end{equation}
\item $\langle\psi\nabla^{2}\psi\rangle$:\\\\
The Laplacian of $\psi$ is given by
\begin{equation}
\nabla^{2}\psi=\frac{a^{2}}{\sqrt{\gamma}}\partial_{i}\left(\sqrt{\gamma}g^{ij}\partial_{j}\psi\right)=\partial_{\chi}^{2}\psi+\frac{\partial_{\theta}^{2}\psi}{\Sigma^{2}(\chi)}+\frac{\partial_{\varphi}^{2}\psi}{\Sigma^{2}(\chi)\sin^{2}\theta}+\frac{2\Sigma'(\chi)}{\Sigma(\chi)}\partial_{\chi}\psi+\frac{\partial_{\theta}\psi}{\Sigma^{2}(\chi)\tan\theta}\,,
\end{equation}
where $\gamma$ denotes the determinant of the spatial part of the metric. The Fourier transform of $\partial_{i}^{2}\psi$ can be easily deduced by taking the second derivative of (\ref{eq:Fourier psi}) with respect to $x^{i}$:
\begin{equation}
f_{\psi_{,ii}}(\vec{p})=-p_{i}^{2}f_{\psi}(\vec{p})\,.
\end{equation}
Now, by the convolution theorem, the Fourier transform of $\psi\partial_{i}^{2}\psi$ is
\begin{equation}
f_{\psi\psi_{,ii}}(\vec{p})=\int d^{3}q f_{\psi}(\vec{q})f_{\psi_{,ii}}(\vec{p}-\vec{q})=-\int d^{3}q\,(p_{i}-q_{i})^{2}f_{\psi}(\vec{q})f_{\psi}(\vec{p}-\vec{q})\,.
\end{equation}
Therefore, 
\begin{equation}
\langle\psi\partial_{i}^{2}\psi\rangle=-\int d^{3}p\int d^{3}q\,(p_{i}-q_{i})^{2}\langle f_{\psi}(\vec{q})f_{\psi}(\vec{p}-\vec{q})\rangle e^{i\vec{p}\cdot\vec{x}}\,,
\end{equation}
which yields
\begin{equation}
\langle\psi\partial_{i}^{2}\psi\rangle=-\frac{1}{(2\pi)^3}\int d^{3}q\,\frac{q_{i}^{2}\langle|\alpha_{\vec{q}}|^{2}\rangle}{\omega_{\vec{q}}}
\end{equation}
once we use (\ref{eq:power fpsi}) and perform the $d^{3}p$ integral.

Regarding the $\psi\partial_{i}\psi$ terms, using (\ref{eq:f psi comma i}) and the convolution theorem, we get
\begin{equation}
f_{\psi\psi_{,i}}(\vec{p})=\int d^{3}q f_{\psi}(\vec{q})f_{\psi_{,i}}(\vec{p}-\vec{q})=i\int d^{3}q\,(p_{i}-q_{i})f_{\psi}(\vec{q})f_{\psi}(\vec{p}-\vec{q})\,.
\end{equation}
Then, the average $\langle\psi\partial_{i}\psi\rangle$ is
\begin{equation}
\langle\psi\partial_{i}\psi\rangle=-i\int d^{3}q\frac{q_{i}\langle|\alpha_{\vec{q}}|^{2}\rangle}{\omega_{\vec{q}}}\,.
\end{equation}
Putting everything together and taking into account that $\vec{q}^{\,\,2}=a^{2}g^{ij}q_{i}q_{j}$, we obtain
\begin{equation}
\langle\psi\nabla^{2}\psi\rangle=-\frac{1}{(2\pi)^3}\int d^{3}q\,\frac{\vec{q}^{\,\,2}\langle|\alpha_{\vec{q}}|^{2}\rangle}{\omega_{\vec{q}}}\,.
\end{equation}
\item $\langle\psi\psi'\rangle$:\\\\
Application of the convolution theorem and (\ref{eq:power fpsi fpi}) immediately yields
\begin{equation}
\langle\psi\psi'\rangle=\int d^{3}p\int d^{3}q\,\langle f_{\psi}(\vec{q})f_{\pi}(\vec{p}-\vec{q})\rangle e^{i\vec{p}\cdot\vec{x}}=0\,.
\end{equation}

\end{itemize}

\end{document}